\documentclass[a4paper]{article}

\usepackage[english]{babel}
\usepackage[utf8x]{inputenc}
\usepackage[T1]{fontenc}

\usepackage[a4paper,top=3cm,bottom=2cm,left=3cm,right=3cm,marginparwidth=1.75cm]{geometry}

\usepackage{amsmath}
\usepackage{authblk}
\usepackage{bm}
\usepackage{graphicx}
\usepackage{multirow}
\usepackage{caption}  
\usepackage{booktabs}
\usepackage{algorithm}
\usepackage{algpseudocode}
\providecommand{\keywords}[1]{\textbf{\textit{Keywords:}} #1}

\title{Separating common (global and local) and distinct variation in multiple mixed types data sets}

\author{Yipeng Song, Johan A. Westerhuis, Age K. Smilde}

\affil{Swammerdam Institute for Life Sciences, University of Amsterdam}

\date{}

\begin{document}

\maketitle

\begin{abstract}
Multiple sets of measurements on the same objects obtained from different
platforms may reflect partially complementary information of the studied system. The
integrative analysis of such data sets not only provides us with the opportunity of
a deeper understanding of the studied system, but also introduces some new
statistical challenges. First, the separation of information that is common
across all or some of the data sets, and the information that is specific to
each data set is problematic. Furthermore, these data sets are often a mix of
quantitative and discrete (binary or categorical) data types, while commonly used data
fusion methods require all data sets to be quantitative. In this paper, we
propose an exponential family simultaneous component analysis (ESCA) model
to tackle the potential mixed data types problem of multiple data sets. In
addition, a structured sparse pattern of the loading matrix is induced through a
nearly unbiased group concave penalty to disentangle the global, local
common and distinct information of the multiple data sets. A Majorization-Minimization based algorithm is derived to fit the proposed model. Analytic solutions are derived for updating all the parameters of the model in each iteration, and the algorithm will decrease the objective function in each iteration monotonically. For model selection, a missing value based cross validation procedure is implemented. The advantages of the proposed method in comparison with other approaches are assessed using comprehensive simulations as well as the analysis of real data from a chronic lymphocytic leukaemia (CLL) study.\\

\noindent Availability: the codes to reproduce the results in this article are available at https://gitlab.com/uvabda.\\

\end{abstract}


\section{Introduction}
Multiple data sets measured on the same samples are becoming increasingly common in different research areas, from biology, food science to psychology. One typical example from biological research is the GDSC1000 study, in which 926 cell lines are fully characterized with respect to point mutation, copy number alternation (CNA), methylation, gene expression and drug responses \cite{iorio2016landscape}. However, these comprehensive measurements from the same cell lines not only provide the opportunity for a deeper understanding of the studied biological system, but also introduce statistical challenges.\\

The first challenge is how to separate the information that is common across all or some of the data sets, and the information which is specific to each data set (often called distinct). These different sources of information have to be disentangled from every data set to have a holistic understanding of the studied system. The second challenge is that measurements from different platforms can be of different data type, such as binary, quantitative or counts. These different data types have different mathematical properties, which should be taken into account in the data analysis. For example, the measurement of a binary variable only has two possible exclusive outcomes, often classified as ``1'' and ``0''. Examples of binary data in biology include point mutation, and the binarized CNA and methylation data sets \cite{iorio2016landscape}. Taking binary measurements ``1'', ``0'' as the quantitative values 1, 0, and casting them into the classical data fusion methods that assume data sets to be quantitative, clearly neglects their binary nature.\\

In this paper, we focus on the component or latent variable based data fusion approaches, although other approaches exist such as undirected graphical model based methods which are able to explore the association between data sets of different data types \cite{aben2018itop}, or between variables of different data types \cite{lee2015learning,cheng2017high}. Two commonly used latent variable based data fusion methods are simultaneous component analysis (SCA) \cite{van2009structured} and iCluster \cite{shen2009integrative}, which both focus on using low dimensional structures to approximate the common variation across all the data sets. Both of these approaches have already been generalized to discrete data sets \cite{mo2013pattern,song2018generalized}. In addition, the concept of common and distinct variation in data fusion has been framed in \cite{van2016separating,smilde2017common}, and several methods \cite{lock2013joint,lofstedt2013global,schouteden2014performing,maage2012preference,smilde2017common} have been proposed. One typical example is JIVE \cite{lock2013joint}. The JIVE model directly specifies the components for the global common variation (variation across all the data sets) and the distinct variation (variation specific to each data set) in the model, and estimates them simultaneously. However, in the JIVE model the local common variation (variation across some of the data sets) is ignored. Direct generalization of JIVE to account for the local common variation is infeasible as with the increase of the number of data sets, the possible combinations of local common variation blows up exponentially. Other methods \cite{schouteden2014performing,lofstedt2013global} encounter similar problems with respect to the estimation of local common variation. In addition, the model selection procedure in these methods is still an unsolved issue \cite{maage2018performance}. A promising solution was provided in \cite{klami2015group,gaynanova2017structural}, in which a group regularization procedure was applied to provide structured sparsity on the loading matrix where the loadings of all variables of a given data set are forced to 0 to disentangle the common (global and local) and distinct variation indirectly. Details will be shown in the following model section. In the SLIDE model \cite{gaynanova2017structural}, first a series of structured sparsity patterns on the loading matrix of a SCA model are learned using a group lasso penalty. Then, these learned structured sparsity patterns are imposed as hard constraints on the loading matrix of a SCA model, and the appropriate model is selected by Bi-cross-validation \cite{perry2009cross}. The Bayesian counterpart of the SLIDE model is the group factor analysis model \cite{klami2015group}, and the generalization of the group factor analysis to mixed data types is the MOFA model \cite{argelaguet2018multi}. These two Bayesian models use automatic relevance determination procedure to induce the structured sparsity.\\

The first contribution in this paper is the generalization of the SCA model to the exponential family SCA (ESCA) model by exploiting the exponential family distribution to account for potentially different data types, such as binary, quantitative or count data. The generalization is done in a similar way as the extension of principal component analysis (PCA) to exponential family PCA \cite{collins2002generalization}. The second contribution is the use of a nearly unbiased group concave penalty to induce a structured sparse pattern on the loading matrix of the ESCA model to disentangle the common (global and local) and distinct variation of multiple data sets of mixed data types. In the SLIDE model \cite{gaynanova2017structural}, the structured sparse pattern is induced by the group lasso penalty, which shrinks in the group level (the groups) as a lasso penalty, and in the individual level (the individual elements inside a group) as a ridge regression penalty. However, a lasso type penalty leads to biased parameter estimation, as the same degree of shrinkage is applied to all the parameters. This will shrink the nonzero parameters too much and as a result makes the prediction or cross validation error based model selection procedures inconsistent \cite{meinshausen2006high,leng2006note}. On the other hand, concave penalties, such as generalized double Pareto (GDP) shrinkage \cite{armagan2013generalized} or bridge ($L_{q: 0<q \leq 1}$) penalty \cite{fu1998penalized}, are capable to achieve nearly unbiased estimation of the parameters while producing sparse solutions. Therefore, we replaced the group lasso penalty by a group concave penalty on the loadings. The group concave penalty shrinks the group level as a concave penalty, and it shrinks on the individual level as ridge regression penalty. The third contribution lies in the derived model fitting algorithm and the model selection procedure. A Majorization-Minimization based algorithm is derived to fit the proposed penalized ESCA (P-ESCA) model. Analytical form solutions for updating all the parameters of the model in each iteration are derived, and the algorithm will decrease the objective function in each iteration monotonically. Furthermore, the missing value problem is tackled in the developed algorithm, and this option is used in the cross validation procedure for the model selection. The proposed model is similar to the MOFA model, but differences exist in the way how the model is derived, how the structured sparsity is achieved, and how the model is selected.\\

Both the performance of the proposed P-ESCA model, and the effectiveness of the model selection procedure are validated by extensive simulations under different situations. The performance of the P-ESCA method is compared with SLIDE and MOFAs. Finally, P-ESCA is exemplified by the explorative analysis of the chronic lymphocytic leukaemia (CLL) data sets \cite{dietrich2018drug,argelaguet2018multi}.\\

\section{P-ESCA model}
In this section, we will introduce the generalization of the ESCA model. And, we will show how to use the group concave penalty to induce the structured sparse pattern on the loading matrix of an ESCA model to disentangle the common (global and local) variation and distinct variation of multiple data sets.\\

\subsection{Exponential family SCA}
The quantitative measurements from $L$ different platforms on the same $I$ objects result into $L$ quantitative data sets, $\left\{\mathbf{X}_l \right\}_{l=1}^{L}$, and the $l^{\text{th}}$ data set $\mathbf{X}_l$($I \times J_l$) has $J_l$ variables. In the classical SCA model, we decompose the $L$ data sets as $\mathbf{X}_l = \mathbf{1}\bm{\mu}_l^{\text{T}} + \mathbf{AB}_l^{\text{T}} + \mathbf{E}_l$, in which $\mathbf{1}$($I\times 1$) is a column vector with ones; $\bm{\mu}_l$($J_l \times 1$) is the column offset term; $\mathbf{A}$($I\times R$) is the common score matrix; $\mathbf{B}_l$($J_l\times R$) and $\mathbf{E}_l$($I\times J_l$) are the loading matrix and residual term respectively for $\mathbf{X}_l$ and $R$ is the number of components. In addition, constraints $\mathbf{A}^{\text{T}}\mathbf{A} = \mathbf{I}$ and $\mathbf{1}^{\text{T}}\mathbf{A} = \mathbf{0}$, in which $\mathbf{I}$ is the identity matrix, are imposed to make the model identifiable. The SCA model tries to discover the common column subspace, which is spanned by the columns of the score matrix $\mathbf{A}$, in $L$ data sets to represent their common information. The column offset terms $\left\{ \bm{\mu}_l \right\}_{l=1}^{L}$ can be removed by column centering of the corresponding data sets $\left\{\mathbf{X}_l\right\}_{l=1}^{L}$. The parameters in the SCA model can be estimated by minimizing the sum of squares $\sum_{l}^{L} w_l ||\mathbf{X}_l - \mathbf{1}\bm{\mu}_l^{\text{T}} - \mathbf{AB}_l^{\text{T}}||_F^2$, in which $w_l$ is the relative weight of the $l^{\text{th}}$ data set $\mathbf{X}_l$.\\

The least squares loss criterion in the classical SCA model is only appropriate for quantitative data sets. When some or all data sets are of another data type, such as binary data, classical SCA model is not appropriate anymore. Motivated by the previous research on exponential family PCA model \cite{schein2003generalized}, we use the exponential family distribution to account for the different data types of multiple data sets, such as Bernoulli distribution for binary data, Poisson distribution for count data and Gaussian distribution for quantitative data.\\

Assume $x \in \mathbf{R}$ follows the exponential dispersion family distribution \cite{agresti2013categorical}, and $\theta$ and $\alpha$ are the natural parameter and the dispersion parameter respectively. The probability density or mass function can be specified as $p(x|\theta,\alpha) = \exp \left[(x\theta - b(\theta))/\alpha \right] h(x,\alpha)$, in which $b(\theta)$ is the log-partition function, and $h(x,\alpha)$ is a function which does not depend on the natural parameter $\theta$. Tab.~S1 lists the log-partition function $b(\theta)$ and its first and second order derivatives $b^{'}(\theta)$, $b^{''}(\theta)$ for Gaussian, Bernoulli and Poisson distributions. The relationship $\text{E}(x|\theta) = b^{'}(\theta)$ always hold in the exponential family distribution. Fig.~S1 visualizes this relationship for the Gaussian, Bernoulli and Poisson distributions. If the $l^{\text{th}}$ data set $\mathbf{X}_l$ is quantitative, according to the probabilistic interpretation of the PCA model \cite{tipping1999probabilistic}, we assume there is a natural parameter matrix $\mathbf{\Theta}_l$($I\times J_l$) underlying $\mathbf{X}_l$, and the low dimensional structure exists in $\mathbf{\Theta}_l$, $\mathbf{X}_l = \mathbf{\Theta}_l + \mathbf{E}_l$ and $\mathbf{\Theta}_l = \mathbf{1}\bm{\mu}_l^{\text{T}} + \mathbf{A}_l\mathbf{B}_l^{\text{T}}$, and elements in the error term $\mathbf{E}_l$ follows a normal distribution $\epsilon_{ij}^{l} \sim \text{N}(0,\sigma^2)$. The conditional mean of the observed $\mathbf{X}_l$ given the low dimensional structure assumption is $\text{E}(\mathbf{X}_l| \mathbf{\Theta}_l) = b^{'}(\mathbf{\Theta}) = \mathbf{\Theta}_l$, in which $b^{'}()$ is the first order derivative of the log-partition function for the Gaussian distribution (Tab.~S1). In exponential family PCA, the same idea has been generalized to other members of exponential family distributions by assuming $\text{E}(\mathbf{X}_l| \mathbf{\Theta}_l) = b^{'}(\mathbf{\Theta}_l)$ and $\mathbf{\Theta}_l = \mathbf{1}\bm{\mu}_l^{\text{T}} + \mathbf{A}_l\mathbf{B}_l^{\text{T}}$, in which the function form of $b^{'}()$ depends on the used probability distribution (Tab.~S1).\\

In the exponential family PCA model, the elements in $\mathbf{X}_l$ are conditionally independent, given the low dimensional structure assumption as $\mathbf{\Theta}_l = \mathbf{1}\bm{\mu}_l^{\text{T}} + \mathbf{A}_l\mathbf{B}_l^{\text{T}}$. Take $x_{ij}^{l}$ and $\theta_{ij}^{l}$ as the $ij^{\text{th}}$ element of $\mathbf{X}_l$ and $\mathbf{\Theta}_l$ respectively. The conditional log-likelihood of observing $\mathbf{X}_l$ can be expressed as $\log(p(\mathbf{X}_l|\mathbf{\Theta}_l,\alpha_l)) =  \sum_{i}^{I}\sum_{j}^{J_l} \log(p(x_{ij}^{l}|\theta_{ij}^{l},\alpha_l)) = \sum_{i}^{I}\sum_{j}^{J_l} \frac{1}{\alpha_l}  (x_{ij}^{l}\theta_{ij}^{l} - b_l(\theta_{ij}^{l})) + c = \frac{1}{\alpha_l} \left[<\mathbf{X}_{l},\mathbf{\Theta}_l> - <\mathbf{1}\mathbf{1}^{\text{T}}, b_l(\mathbf{\Theta}_l)>\right] + c$, in which $<,>$ indicates the inner product, for matrices, $<\mathbf{X}_l,\mathbf{\Theta}_l> = \text{trace}(\mathbf{X}_l^{\text{T}}\mathbf{\Theta}_l)$; $c$, a constant does not depend on the unknown parameter $\mathbf{\Theta}_l$; $b_l()$ and $\alpha_l$ are the element-wise log-partition function and the known dispersion parameter respectively for the $l^{\text{th}}$ data set $\mathbf{X}_l$. In the ESCA model, we assume that the natural parameter matrices $\left\{ \mathbf{\Theta}_l \right\}_{l=1}^{L}$ lie in the same column subspace, which is spanned by the common score matrix $\mathbf{A}$. To make the model identifiable, constraints $\mathbf{A}^{\text{T}}\mathbf{A} = \mathbf{I}$ and $\mathbf{1}^{\text{T}}\mathbf{A} = \mathbf{0}$ are imposed. The optimization problem associated with this ESCA model can be expressed as follows,
\begin{equation}\label{eq1}
\begin{aligned}
    \min_{ \left\{\bm{\mu}_l\right\}_{l}^{L}, \mathbf{A}, \left\{\mathbf{B}_l\right\}_{l}^{L}} \quad & \sum_{l=1}^{L} -\log(p(\mathbf{X}_l|\mathbf{\Theta}_l, \alpha_l))\\
	& =\sum_{l=1}^{L} \frac{1}{\alpha_l} \left[ <\mathbf{1}\mathbf{1}^{\text{T}}, b_l(\mathbf{\Theta}_l)> - <\mathbf{X}_l, \mathbf{\Theta}_l> \right] + c\\
    \text{s.t.~} \mathbf{\Theta}_l &= \mathbf{1}\bm{\mu}_l^{\text{T}} + \mathbf{AB}_l^{\text{T}}, \quad l = 1,\ldots,L \\
     \mathbf{1}^{\text{T}}\mathbf{A} &= \mathbf{0}\\
	 \mathbf{A}^{\text{T}}\mathbf{A} &= \mathbf{I}.
\end{aligned}
\end{equation}

\subsection{Separating common (global and local) and distinct variation in multiple data sets via structured sparsity}
The drawback of the SCA or ESCA models is that only the global common components, which account for the common variation across all the data sets, is allowed. However, the real situation in multiple data sets integration can be far more complex as local common variation across some of the data sets and distinct variation in each data set are expected as well. Directly specifying the components in the ESCA model for common (global and local) and distinct variation in the same way as JIVE model \cite{lock2013joint} is infeasible, as the number of possible combinations of local common variation will blow up exponentially with an increasing number of data sets. A promising solution is using structured sparsity on the loading matrix to disentangle the common (global and local) and distinct variation indirectly \cite{klami2015group,gaynanova2017structural}. Structured sparsity of the data set specific loading matrices in component based data fusion methods has been explored by \cite{van2011flexible,acar2015data}. The idea of using structured sparsity to disentangle the common (global and local) and distinct variation in multiple quantitative data sets is made explicit in \cite{klami2015group,gaynanova2017structural}. To illustrate the idea, we use an example with three quantitative data sets. Suppose we construct a SCA model on three column centered quantitative data sets $\left\{\mathbf{X}_l \right\}_{l=1}^{3}$, the common score matrix is $\mathbf{A}$, the corresponding loading matrices are $\left\{\mathbf{B}_l \right\}_{l=1}^{3}$, and $\mathbf{X}_l = \mathbf{A}\mathbf{B}_{l}^{\text{T}} + \mathbf{E}_l$, in which $\mathbf{E}_l$ is the residual term for $l^{\text{th}}$ data set. If the structured sparsity pattern in $\left\{\mathbf{B}_l \right\}_{l=1}^{3}$ is expressed as follows,
\begin{equation*}
\begin{aligned}
   \left(
                 \begin{array}{c}
                   \mathbf{B}_1 \\
                   \mathbf{B}_2 \\
                   \mathbf{B}_3 \\
                 \end{array}
               \right)
               = \left(
                   \begin{array}{cccccccc}
                     \mathbf{b}_{1,1} & \mathbf{b}_{1,2} & \mathbf{b}_{1,3} & \mathbf{0}       & \mathbf{b}_{1,5} & \mathbf{0}       & \mathbf{0}      \\
                     \mathbf{b}_{2,1} & \mathbf{b}_{2,2} & \mathbf{0}       & \mathbf{b}_{2,4} & \mathbf{0}       & \mathbf{b}_{2,6} & \mathbf{0}       \\
                     \mathbf{b}_{3,1} & \mathbf{0}       & \mathbf{b}_{3,3} & \mathbf{b}_{3,4} & \mathbf{0}       & \mathbf{0}       & \mathbf{b}_{3,7} \\
                   \end{array}
                 \right),
\end{aligned}
\end{equation*}
in which $\mathbf{b}_{l,r} \in \mathbf{R}^{J_l}$ indicates the $r^{\text{th}}$ column of the $l^{\text{th}}$ loading matrix $\mathbf{B}_l$, then we have the following relationships,
\begin{equation*}
\begin{aligned}
   \mathbf{X}_1 & = \mathbf{a}_1\mathbf{b}_{1,1}^{\text{T}} &+& \mathbf{a}_2\mathbf{b}_{1,2}^{\text{T}} &+& \mathbf{a}_3\mathbf{b}_{1,3}^{\text{T}} &+& \mathbf{0}                     &+& \mathbf{a}_5\mathbf{b}_{1,5}^{\text{T}} &+& \mathbf{0}                     &+& \mathbf{0}  &+& \mathbf{E}_1                   \\
   \mathbf{X}_2 & = \mathbf{a}_1\mathbf{b}_{2,1}^{\text{T}} &+& \mathbf{a}_2\mathbf{b}_{2,2}^{\text{T}} &+& \mathbf{0}                     &+& \mathbf{a}_4\mathbf{b}_{2,4}^{\text{T}} &+& \mathbf{0}                     &+& \mathbf{a}_6\mathbf{b}_{2,6}^{\text{T}} &+& \mathbf{0}  &+& \mathbf{E}_2                   \\
   \mathbf{X}_3 & = \mathbf{a}_1\mathbf{b}_{3,1}^{\text{T}} &+& \mathbf{0}                     &+& \mathbf{a}_3\mathbf{b}_{3,3}^{\text{T}} &+& \mathbf{a}_4\mathbf{b}_{3,4}^{\text{T}} &+& \mathbf{0}                     &+& \mathbf{0}                     &+& \mathbf{a}_7\mathbf{b}_{3,7}^{\text{T}} &+& \mathbf{E}_3 .
\end{aligned}
\end{equation*}
Here $\mathbf{a}_r$ indicates the $r^{\text{th}}$ column of the common score matrix $\mathbf{A}$. The first component represents the global common variation across three data sets; the $2^{\text{nd}}$, $3^{\text{nd}}$ and $4^{\text{nd}}$ components represent the local common variation across two data sets and the $5^{\text{nd}}$, $6^{\text{nd}}$ and $7^{\text{nd}}$ components represent the distinct variation specific to each single data set. In this way, the structured sparsity pattern in the loading matrices $\left\{\mathbf{B}_l \right\}_{l=1}^{3}$ can be used to separate the common (global and local) and distinct variation of multiple quantitative data sets.\\

\subsection{Group concave penalty}
In \cite{van2011flexible,acar2015data,gaynanova2017structural}, the structured sparsity is induced by a group lasso penalty on the columns of $\{ \mathbf{B}_l \}_{1}^{L}$. The used group lasso penalty is $\lambda \sum_{l}\sum_{r} ||\mathbf{b}_{l,r}||_{2}$, in which $\lambda$ is the tuning parameter, $\mathbf{b}_{l,r}$ indicates the $r^{\text{th}}$ column of the $l^{\text{th}}$ loading matrix $\mathbf{B}_l$, and $||\quad||_{2}$ indicates the $L_2$ norm of a vector. This group lasso penalty shrinks $||\mathbf{b}_{l,r}||_{2}$ as a lasso penalty and the elements inside $\mathbf{b}_{l,r}$ as a ridge penalty. However, lasso type penalty leads to biased parameter estimation as the same degree of shrinkage is applied to all the parameters, which will shrink the nonzero parameters too much and makes the prediction or cross validation error based model selection procedures inconsistent \cite{meinshausen2006high,leng2006note}. This leads in general to the selection of too complex models. The SLIDE model \cite{gaynanova2017structural} solves the model selection problem in a two stages manner. First, varying degrees of regularization are imposed to induce a series of structured sparse loading patterns. Then these structured sparse patterns are taken as hard constraints on a new SCA model, in which a Bi-cross validation procedure \cite{perry2009cross} is used for the final selection. This two stages approach is similar to the often used re-estimation trick in lasso regression. However, such a two-step strategy cannot easily be generalized to the ESCA model. For example, if a binary data set is used and the structured sparse pattern is imposed as a hard constraint on the loading matrices in a ESCA model, the estimated loadings of the binary data set can easily go to infinity \cite{groenen2016multinomial,song2018generalized}.\\

The above issue introduced by the biased estimation of lasso type penalties can be alleviated by using concave penalties \cite{fu1998penalized,armagan2013generalized}, which can achieve sparse solutions and nearly unbiased parameter estimation simultaneously. Therefore, in this paper, we applied group concave penalties, generalized double Pareto (GDP) shrinkage \cite{armagan2013generalized} and bridge ($L_{q: 0<q \leq 1}$) penalty \cite{fu1998penalized} are included as special cases, on the loading matrices of the ESCA model to induce structured sparse pattern. Take $\sigma_{lr} = ||\mathbf{b}_{l,r}||_2$, in which $\mathbf{b}_{l,r}$ is the $r^{\text{th}}$ column of $\mathbf{B}_l$, and $g()$ is a general concave penalty function in Tab.~1. The penalty on $\mathbf{B}_l$ can be expressed as $\lambda_l \sum_{r} g(\sigma_{lr})$, in which $\lambda_l$ is the tuning parameter. The group lasso penalty is a special case of the group $L_{q}$ (bridge) penalty by setting $q=1$. The thresholding properties of the group $L_{\text{q}}$ penalty, group GDP penalty and group lasso can be found in Fig.~S2. In order to account for the situation that the data sets have an unequal number of variables, we add the weights in the same way as in the standard group lasso regression problem, i.e. $\lambda_l \sqrt{J_l} \sum_{r} g(\sigma_{lr})$. The group concave penalty on $\{ \mathbf{B}_l \}_{1}^{L}$ can be expressed as $\sum_{l} \Big[\lambda_l \sqrt{J_l} \sum_{r} g(\sigma_{lr}) \Big]$. Based on successful results in previous work \cite{song2018generalized} we will focus on the GDP penalty, which is differentiable everywhere in its domain and its performance is insensitive to the selection of the hyper-parameter $\gamma$. We given an example in Fig.~S3 to show how the group GDP ($\gamma = 1$) penalty induces structured sparsity pattern on the loading matrices $\left\{\mathbf{B} \right\}_{l=1}^3$.\\

\begin{table}[h!]
\centering
\caption*{\textbf{Tab.~1}: Three commonly used group penalty functions. Take $\sigma$ as the $L_2$ norm of a group of elements. $q$ and $\gamma$ are the tuning parameters. The supergradient is the counter concept of the subgradient for a concave function. When the concave function is differentiable everywhere, the supergradient is the gradient.}
\label{Table:1}
\begin{tabular}{lll}
  \toprule
penalty & formula & supergradient \\
  \midrule
 group lasso & $ \sigma $ & $1$ \\

group $L_{q: 0<q \leq 1}$ & $ \sigma^q $ & $\left\{ \begin{array}{ll} +\infty &\textrm{$\sigma=0$}\\
                                 q \sigma^{q-1} &\textrm{$\sigma>0$}\\ \end{array} \right.$ \\
group GDP & $ \log(1+\frac{\sigma}{\gamma}) $ & $\frac{1}{\gamma + \sigma}$ \\
  \bottomrule
\end{tabular}
\end{table}

\subsection{Identifiability}
The constraint $\mathbf{1}^{\text{T}}\mathbf{A} = \mathbf{0}$ makes the column offset terms $\left\{ \bm{\mu} \right\}_{l=1}^{L}$ identifiable. The columns of the score matrix $\mathbf{A}$ span the joint subspace $\bigcup_{l=1}^L \text{col}(\mathbf{\Theta}_l)$, in which $\text{col}()$ indicates the column subspace. The structured sparse pattern on the loading matrices and the multiplication of the score and loading matrices provide a way to separate the joint subspace $\bigcup_{l=1}^L \text{col}(\mathbf{\Theta}_l)$ into subspaces $ \text{col(GC)}$, $\text{col(LC)}$, $\text{col(D)}$ corresponding to the global common, local common and distinct variation, $ \text{col(GC)} \bigcup \text{col(LC)} \bigcup \text{col(D)} = \bigcup_{l=1}^L \text{col}(\mathbf{\Theta}_l)$. If the orthogonality constraint $\mathbf{A}^{\text{T}}\mathbf{A} = \mathbf{I}$ is imposed, the separated subspaces $ \text{col(GC)}$, $\text{col(LC)}$, $\text{col(D)}$, corresponding to the global common, local common and distinct variation, are orthogonal to each other, and unique as $\text{col(GC)} \bigcap \text{col(LC)} \bigcap \text{col(D)} = \emptyset$. Furthermore, there is no rotational freedom for the components within the subspace corresponding to the global common or local common or distinct variation. This is because an orthogonal rotation operation will alter the value of the penalty function on the loading matrix even though the structured sparse pattern is unchanged. Since the separated subspaces are unique and there is no rotational freedom for the subspaces, the score matrix $\mathbf{A}$ and loading matrix $\mathbf{B}$ are unique. Therefore, the model is identifiable with respect to the parameters $\bm{\mu}$, $\mathbf{A}$ and $\mathbf{B}$.\\

\subsection{Regularized likelihood criterion}
The regularized likelihood criterion of fitting the proposed P-ESCA model can be derived as follows. To tackle the missing value problem, $L$ weighting matrices are introduced. For the $l^{\text{th}}$ data set $\mathbf{X}_l$, we introduce a same size weighting matrix $\mathbf{W}_l$, in which $w_{ij}^{l} = 0$ if the corresponding element in $\mathbf{X}_l$ is missing, while $w_{ij}^{l} = 1$  \textit{vise versa}. This option is the basis for different missing value based cross validation approaches. The corresponding optimization problem can be expressed as follows,
\begin{equation}\label{eq2}
\begin{aligned}
    \min_{ \left\{\bm{\mu}_l\right\}_{l}^{L}, \mathbf{A}, \left\{\mathbf{B}_l\right\}_{l}^{L}} \quad & \sum_{l=1}^{L} \Big[ -\log(p(\mathbf{X}_l|\mathbf{\Theta}_l, \alpha_l)) + \lambda_l \sqrt{J_l} \sum_{r} g(\sigma_{lr}) \Big] \\
	&= \sum_{l=1}^{L} \Big[ \frac{1}{\alpha_l}( <\mathbf{W}_l, b_l(\mathbf{\Theta}_l)> - <\mathbf{W}_l \odot \mathbf{X}_l,\mathbf{\Theta}_l>) + \lambda_l \sqrt{J_l} \sum_{r} g(\sigma_{lr}) \Big] + c\\
    \text{s.t.~} \mathbf{\Theta}_l &= \mathbf{1}\bm{\mu}_l^{\text{T}} + \mathbf{AB}_l^{\text{T}}, \quad l = 1,\ldots,L \\
     \mathbf{1}^{\text{T}}\mathbf{A} &= \mathbf{0}\\
	 \mathbf{A}^{\text{T}}\mathbf{A} &= \mathbf{I} \\
	 \sigma_{lr} &= ||\mathbf{b}_{l,r}||_2, l = 1...L; r = 1,\ldots, R,
\end{aligned}
\end{equation}
in which $\odot$ indicates the element-wise matrix multiplication.\\

\section{Algorithm}
The original problem in equation \ref{eq2} is difficult to optimize directly because of the non-convex orthogonality constraint $\mathbf{A}^{\text{T}}\mathbf{A} = \mathbf{I}$ and the group concave penalty $g()$. However, by using the Majorization-Minimization (MM) principle, the original difficult problem can be majorized to a simpler problem, for which analytical form solutions can be derived for all the parameters. According to the MM principal, the derived algorithm will monotonously decrease the loss function in each iteration. Further details of the MM principle can be found in \cite{de1994block,hunter2004tutorial}.\\

\subsection{The majorization of the regularized likelihood criterion}
Take $f_l(\mathbf{\Theta}_l) = \frac{1}{\alpha_l} \left[<\mathbf{W}_l, b_l(\mathbf{\Theta}_l)> - <\mathbf{W}_l \odot \mathbf{X}_l, \mathbf{\Theta}_l>\right]$ as the loss function for fitting the $l^{\text{th}}$ data set $\mathbf{X}_l$, and $g_l(\mathbf{B}_l) = \sum_{r} g(\sigma_{lr})$ as the group concave penalty for the $l^{\text{th}}$ loading matrix $\mathbf{B}_l$. We can majorize $f_l(\mathbf{\Theta}_l)$ and $g_l(\mathbf{B}_l)$ respectively as follows.\\

\subsubsection*{The majorization of $f_l(\mathbf{\Theta}_l)$}
Given $\tilde{f_l}(\theta_{ij}^{l}) = b_l(\theta_{ij}^{l}) - x_{ij}^{l}\theta_{ij}^{l}$, we have $f_l(\mathbf{\Theta}_l) = \frac{1}{\alpha_l} \sum_{i}\sum_{j} w_{ij}^{l} \tilde{f_l}(\theta_{ij}^{l})$. The first and second gradients of $\tilde{f_l}(\theta_{ij}^{l})$ with respect to $\theta_{ij}^{l}$ are $\nabla \tilde{f_l}(\theta_{ij}^{l}) = b_l^{'}(\theta_{ij}^{l}) - x_{ij}^{l}$ and $\nabla^2 \tilde{f_l}(\theta_{ij}^{l}) = b_l^{''}(\theta_{ij}^{l})$. Assume that $\nabla^2 \tilde{f_{1}}(\theta_{ij}^{l})$ is upper bounded by a constant $\rho_{l}$, which will be detailed below. If $\theta^{l}$ represents the general representation of $\theta_{ij}^{l}$, then according to the Taylor's theorem and the assumption that $\nabla^2 \tilde{f_{l}}(\theta^{l}) \leq \rho_{l}$ for all $\theta^{l} \in \text{domain}(\tilde{f_l})$, we have the following inequality,
\begin{equation}\label{eq3}
\begin{aligned}
\tilde{f_l}(\theta^{l}) &= \tilde{f_l}((\theta^{l})^k) + <\nabla \tilde{f_l}((\theta^{l})^k), \theta^{l}-(\theta^{l})^k> + \frac{1}{2}(\theta^{l}-(\theta^{l})^k)^{\text{T}} \nabla^{2} \tilde{f_l}\left[(\theta^{l})^k + t(\theta^{l}-(\theta^{l})^k)\right](\theta^{l}-(\theta^{l})^k) \\
          &\leq \tilde{f_l}((\theta^{l})^k) + <\nabla \tilde{f_l}((\theta^{l})^k), \theta^{l}-(\theta^{l})^k> + \frac{\rho_{l}}{2}(\theta^{l}-(\theta^{l})^k)^2 \\
          &= \frac{\rho_{l}}{2}\left[\theta^{l}-(\theta^{l})^k + \frac{1}{\rho_l}\nabla \tilde{f_l}((\theta^{l})^k)\right]^2 + c.\\
\end{aligned}
\end{equation}
Here $(\theta^{l})^k$ is an approximation of $\theta^{l}$ at the $k^{\text{th}}$ iteration and $t\in[0,1]$ is an unknown constant. Combining the above inequality and the majorization step \cite{kiers1997weighted} of transforming a weighted least square problem to a least squares problem, we have the following inequality,
\begin{equation}\label{eq4}
\begin{aligned}
     f_l(\mathbf{\Theta}_l) &= \frac{1}{\alpha_l} \sum_{i}\sum_{j} w_{ij}^{l} \tilde{f_l}(\theta_{ij}^{l})\\
     &\leq \frac{\rho_l}{2\alpha_l} ||\mathbf{W}_l \odot (\mathbf{\Theta}_l - \mathbf{\Theta}_l^k + \frac{1}{\rho_l}(b_l^{'}(\mathbf{\Theta}_{l}^{k}) - \mathbf{X}_l)) ||_F^2 + c\\
     &\leq 	\frac{\rho_l}{2\alpha_l} ||\mathbf{\Theta}_l - \mathbf{H}_{l}^{k}||_F^2 + c \\
\mathbf{H}_{l}^{k} &= \mathbf{W}_l \odot(\mathbf{\Theta}_l^k - \frac{1}{\rho_l} (b_l^{'}(\mathbf{\Theta}_{l}^{k}) - \mathbf{X}_l)) + (\mathbf{1}\mathbf{1}^{\text{T}}-\mathbf{W}_l)\odot \mathbf{\Theta}_l^k \\
&= \mathbf{\Theta}_l^k - \frac{1}{\rho_l} \mathbf{W}_l \odot (b_l^{'}(\mathbf{\Theta}_{l}^{k}) - \mathbf{X}_l)), 	
\end{aligned}
\end{equation}
in which $\mathbf{\Theta}_l^k$ is the approximation of $\mathbf{\Theta}_l$ during the $k^{\text{th}}$ iteration. For the Bernoulli distribution, an elegant bound $b^{''}(\theta) \leq 0.25$ can be used \cite{de2006principal}; for the Gaussian likelihood, $b^{''}(\theta) = 1$; for the Poisson distribution, $b^{''}(\theta)$ is unbounded, however, we can always set $\rho_l = \text{max}(b^{''}(\mathbf{\Theta}_l^k))$.\\

\subsubsection*{The majorization of $g_l(\mathbf{B}_l)$}
Assume $\mathbf{B}_l^k$ is the $k^{\text{th}}$ approximation of $\mathbf{B}_l$, and $\sigma_{lr}^k = ||\mathbf{b}_{l,r}^k||_2$. According to the definition of a concave function \cite{boyd2004convex}, we always have the inequality $g(\sigma_{lr}) \leq g(\sigma_{lr}^k) + \omega_{lr}^k(\sigma_{lr} - \sigma_{lr}^k) = \omega_{lr}^k \sigma_{lr} + c$, in which $\omega_{lr}^k \in \partial g(\sigma_{lr}^k)$ and $\partial g(\sigma_{lr}^k)$ is the set of supergradients (the counterpart concept of the subgradient for a concave function) of the function $g()$ at $\sigma_{lr}^k$. When the supergradient is unique, then $\omega_{lr}^k = \partial g(\sigma_{lr}^k)$. Therefore, $g_l(\mathbf{B}_l) = \sum_{r} g(\sigma_{lr})$ can be majorized as follows,
\begin{equation}\label{eq5}
\begin{aligned}
g_l(\mathbf{B}_l) &= \sum_{r} g(\sigma_{lr})\\
              &\leq \sum_{r} \omega_{lr}^k \sigma_{lr} + c\\
       \omega_{lr}^k &\in \partial g(\sigma_{lr}^k).
\end{aligned}
\end{equation}

\subsubsection*{The majorization of the regularized likelihood criterion}
Combining the above two majorization steps, we have majorized the original complex problem in the equation \ref{eq2} to a simper problem in each iteration as follows,
\begin{equation}\label{eq6}
\begin{aligned}
    \min_{ \left\{\bm{\mu}_l\right\}_{l}^{L}, \mathbf{A}, \left\{\mathbf{B}_l\right\}_{l}^{L}} \quad & \sum_{l=1}^{L}\Big[ \frac{\rho_l}{2\alpha_l} ||\mathbf{\Theta}_l - \mathbf{H}_{l}^{k}||_F^2  + \lambda_l \sqrt{J_l} \sum_{r} \omega_{lr}^k \sigma_{lr} \Big] \\
    \text{s.t.~} \mathbf{\Theta}_l &= \mathbf{1}\bm{\mu}_l^{\text{T}} + \mathbf{AB}_l^{\text{T}}, l = 1 \ldots L \\
     \mathbf{1}^{\text{T}}\mathbf{A} &= \mathbf{0}\\
	 \mathbf{A}^{\text{T}}\mathbf{A} &= \mathbf{I} \\
	 \sigma_{lr} &= ||\mathbf{b}_{l,r}||_2, l = 1 \ldots L, r = 1 \ldots R\\
\mathbf{H}_{l}^{k} &= \mathbf{\Theta}_l^k - \frac{1}{\rho_l} \mathbf{W}_l \odot (b_l^{'}(\mathbf{\Theta}_{l}^{k}) - \mathbf{X}_l), l = 1 \ldots L \\
\omega_{lr}^k &\in \partial g(\sigma_{lr}^k), l = 1 \ldots L, r = 1 \ldots R.
\end{aligned}
\end{equation}

\subsection{Block coordinate descent}
The majorized optimization problem in equation \ref{eq6} can be solved by the block coordinate descent approach, and the analytic solution can be derived for all the parameters.\\

\subsubsection*{Updating $\{ \bm{\mu}_l \}_{1}^{L}$}
When fixing all other parameters except $\bm{\mu}_l$, the analytic solution of $\bm{\mu}_l$ in equation (\ref{eq6}) is simply the column mean of $\mathbf{H}_l^k$, $\bm{\mu}_l = \frac{1}{I} (\mathbf{H}_l^k)^{\text{T}} \mathbf{1}$.\\

\subsubsection*{Updating $\mathbf{A}$}
When fixing all other parameters except $\mathbf{A}$, and deflating the offset term $\{ \bm{\mu}_l \}_{1}^{L}$, the loss function in equation (\ref{eq6}) becomes $\sum_{l=1}^{L} \frac{\rho_l}{2\alpha_l} ||\mathbf{AB}_l^{\text{T}} - \mathbf{JH}_{l}^{k}||_F^2 + c$, in which $\mathbf{J} = \mathbf{I} - \frac{1}{I} \mathbf{1} \mathbf{1}^{\text{T}}$ is the column centering matrix. If we take $d_l = \sqrt{\rho_l /\alpha_l}$, the above equation can also be written in this way $\sum_{l=1}^{L} \frac{1}{2} ||\mathbf{A}d_l\mathbf{B}_l^{\text{T}} - d_l\mathbf{JH}_{l}^{k}||_F^2$. To simplify the equations, we set $\widetilde{\mathbf{B}}_l = d_l \mathbf{B}_l$ and $\widetilde{\mathbf{JH}}_l^k = d_l \mathbf{JH}_{l}^{k}$. Then, we take $\widetilde{\mathbf{B}}$ as the row concatenation of $\left\{ \widetilde{\mathbf{B}}_l \right\}_{l=1}^{L}$, $\widetilde{\mathbf{B}}^{\text{T}} = [\widetilde{\mathbf{B}}_1^{\text{T}}  \ldots \widetilde{\mathbf{B}}_l^{\text{T}} \ldots \widetilde{\mathbf{B}}_L^{\text{T}}]$, and take $\widetilde{\mathbf{JH}}^k$ as the column concatenation of $\left\{\widetilde{\mathbf{JH}}_l^k \right\}_{l=1}^{L}$, $\widetilde{\mathbf{JH}}^k = [\widetilde{\mathbf{JH}}_1^k \ldots \widetilde{\mathbf{JH}}_l^k \ldots \widetilde{\mathbf{JH}}_L^k]$. After that, we have $\sum_{l=1}^{L} \frac{\rho_l}{2\alpha_l} ||\mathbf{AB}_l^{\text{T}} - \mathbf{JH}_{l}^{k}||_F^2 = \sum_{l=1}^{L} \frac{1}{2} ||\mathbf{A}\widetilde{\mathbf{B}}_l^{\text{T}} - \widetilde{\mathbf{JH}}_{l}^{k}||_F^2 = \frac{1}{2}||\mathbf{A}\widetilde{\mathbf{B}}^{\text{T}} - \widetilde{\mathbf{JH}}^{k}||_F^2$. Updating $\mathbf{A}$ equivalents to minimizing $\frac{1}{2}||\mathbf{A}\widetilde{\mathbf{B}}^{\text{T}} - \widetilde{\mathbf{JH}}^{k}||_F^2, \text{s.t.} \mathbf{A}^{\text{T}}\mathbf{A} = \mathbf{I}$. Assume the SVD decomposition of $\widetilde{\mathbf{JH}}^{k}\widetilde{\mathbf{B}}$ is $\widetilde{\mathbf{JH}}^{k}\widetilde{\mathbf{B}} = \mathbf{UDV}^{\text{T}}$, the analytic solution for $\mathbf{A}$ is $\mathbf{A} = \mathbf{UV}^{\text{T}}$. The derivation of the above solution is shown in the following paragraph.\\

To simplify the derivation, we take $\mathbf{B} = \widetilde{\mathbf{B}}$ and $\mathbf{C} = \widetilde{\mathbf{JH}}^{k}$. So the optimization problem is $\min_{\mathbf{A}} ||\mathbf{A}\mathbf{B}^{\text{T}} - \mathbf{C}||_F^2, \text{s.t.} \mathbf{A}^{\text{T}} \mathbf{A} = \mathbf{I}$. This equation can be expanded as $||\mathbf{A}\mathbf{B}^{\text{T}} - \mathbf{C}||_F^2 = \text{tr}(\mathbf{BA}^{\text{T}}\mathbf{A}\mathbf{B}^{\text{T}}) - 2\text{tr}(\mathbf{B}\mathbf{A}^{\text{T}}\mathbf{C}) + \text{tr}(\mathbf{C}^{\text{T}}\mathbf{C})$. Since $\mathbf{A}^{\text{T}}\mathbf{A} = \mathbf{I}$, the above optimization problem equivalents to maximizing a trace function problem, $\max_{\mathbf{A}} \text{tr}(\mathbf{B}\mathbf{A}^{\text{T}}\mathbf{C}), \text{s.t.} \mathbf{A}^{\text{T}} \mathbf{A} = \mathbf{I}$. Assume the SVD decomposition of $\mathbf{C}\mathbf{B}$ is $\mathbf{C}\mathbf{B} = \mathbf{UDV}^{\text{T}}$, we have $\text{tr}(\mathbf{B}\mathbf{A}^{\text{T}}\mathbf{C}) = \text{tr}(\mathbf{A}^{\text{T}}\mathbf{C}\mathbf{B}) = \text{tr}(\mathbf{A}^{\text{T}}\mathbf{UDV}^{\text{T}}) = \text{tr}(\mathbf{V}^{\text{T}}\mathbf{A}^{\text{T}}\mathbf{UD})$. According to the Kristof theorem \cite{ten1993least}, we have $\text{tr}(\mathbf{V}^{\text{T}}\mathbf{A}^{\text{T}}\mathbf{UD}) \leq \sum_{r}d_{rr}$, in which $d_{rr}$ is the $r^{\text{th}}$ diagonal element of $\mathbf{D}$, and this upper-bound can be achieved by setting $\mathbf{A} = \mathbf{UV}^{\text{T}}$.\\

\subsubsection*{Updating $\{ \mathbf{B}_l \}_{1}^{L}$}
Because $\mathbf{A}^{\text{T}}\mathbf{A} = \mathbf{I}$, it is easy to prove that $||\mathbf{AB}_l^{\text{T}} - \mathbf{JH}_{l}^{k}||_F^2 = ||\mathbf{A}^{\text{T}}\mathbf{AB}_l^{\text{T}} - \mathbf{A}^{\text{T}}\mathbf{JH}_{l}^{k}||_F^2 = ||\mathbf{B}_l - (\mathbf{JH}_{l}^{k})^{\text{T}} \mathbf{A}||_F^2$. Also, because of that the least squares problems are decomposable, we have $||\mathbf{B}_l - (\mathbf{JH}_{l}^{k})^{\text{T}} \mathbf{A}||_F^2 = \sum_{r} ||\mathbf{b}_{l,r} - (\mathbf{JH}_{l}^{k})^{\text{T}} \mathbf{a}_r||_2^2$, in which $\mathbf{a}_r$ is the $r^{\text{th}}$ column of $\mathbf{A}$. In this way, we have the following optimization problem,
\begin{equation}\label{eq7}
\begin{aligned}
    \min_{\mathbf{B}_l} \quad & \frac{\rho_l}{2\alpha_l} ||\mathbf{AB}_l^{\text{T}} - \mathbf{JH}_{l}^{k}||_F^2  + \lambda_l \sqrt{J_l} \sum_{r} \omega_{lr}^k \sigma_{lr} \\
	&= \frac{\rho_l}{2\alpha_l}||\mathbf{B}_l - (\mathbf{JH}_{l}^{k})^{\text{T}} \mathbf{A}||_F^2 + \lambda_l \sqrt{J_l} \sum_{r} \omega_{lr}^k \sigma_{lr}\\
    &= \sum_{r} \Big[ \frac{\rho_l}{2\alpha_l}(\mathbf{b}_{l,r} - (\mathbf{JH}_{l}^{k})^{\text{T}} \mathbf{a}_r)^2 + \lambda_l \sqrt{J_l}\omega_{lr}^k \sigma_{lr} \Big]\\
    \text{s.t.} \quad \sigma_{lr} &= ||\mathbf{b}_{l,r}||_2, l = 1 \ldots L, r = 1 \ldots R,
\end{aligned}
\end{equation}
The above optimization problem is equivalent to finding the proximal operator of a $L_2$ (or Euclidean) norm, and the analytic solution exists \cite{parikh2014proximal}. Take $\tilde{\lambda}_{lr} = \lambda_l \sqrt{J_l} \omega_{lr}^k \alpha_l/\rho_l$, the analytical solution of $\mathbf{b}_{l,r}$ is $\mathbf{b}_{l,r} = \max(0, 1- \frac{\tilde{\lambda}_{lr}}{||(\mathbf{JH}_{l}^{k})^{\text{T}} \mathbf{a}_r||_2}) (\mathbf{JH}_{l}^{k})^{\text{T}} \mathbf{a}_r$. To update the parameter $\mathbf{B}_l$, we can simply apply this proximal operator to all the columns of $\mathbf{B}_l$.\\

\subsubsection*{Initialization and stopping criteria}
The initialization of the parameters $\left\{\bm{\mu}_l^0\right\}_{l=1}^{L}$, $\mathbf{A}^0$, $\left\{\mathbf{B}_l^0\right\}_{l=1}^{L}$ can be set to the results of a classical SCA model on $\left\{\mathbf{X}_l\right\}_{l=1}^{L}$ or to accept user imputed initializations. The relative change of the objective function is used as the stopping criteria. Pseudocode of the algorithm described above is shown in Algorithm \ref{alg1}, in which $f^k$ is the value of the objective function in $k^{\text{th}}$ iteration, $\epsilon_f$ is the tolerance of relative change of the objective function.\\

\begin{algorithm}[htb]
  \caption{An MM algorithm for fitting the P-ESCA model.}
  \label{alg1}
  \begin{algorithmic}[1]
    \Require
      $\left\{\mathbf{X}_l\right\}_{l=1}^{L}$, $\left\{\alpha_l\right\}_{l=1}^{L}$, $g()$, $\left\{\lambda_l\right\}_{l=1}^{L}$, $\gamma$;
    \Ensure
      $\hat{\bm{\mu}}$, $\hat{\mathbf{A}}$, $\hat{\mathbf{B}}$;
    \State Compute $\left\{\mathbf{W}_l\right\}_{l=1}^{L}$ for missing values in $\left\{\mathbf{X}_l\right\}_{l=1}^{L}$;
    \State Initialize $\left\{\bm{\mu}_l^0\right\}_{l=1}^{L}$, $\mathbf{A}^0$, $\left\{\mathbf{B}_l^0\right\}_{l=1}^{L}$;
    \State $\mathbf{\Theta}_l^0 = \mathbf{1}(\bm{\mu}_l^0)^{\text{T}} + \mathbf{A}^0(\mathbf{B}_l^0)^{\text{T}}, l = 1 \ldots L$;
    \State $k = 0$;
    \While{$(f^{k-1}-f^{k})/f^{k-1}>\epsilon_f$}
        \For{$l=1\ldots L$}
            \State \text{Estimate} $\rho_l$ \text{according to the data type of} $\mathbf{X}_l$;
            \State  $\mathbf{H}_{l}^{k} = \mathbf{\Theta}_l^k - \frac{1}{\rho_l} \mathbf{W}_l \odot (b^{'}(\mathbf{\Theta}_{l}^{k}) - \mathbf{X}_l))$;
            \State $\bm{\mu}_l^{k+1} = \frac{1}{I} (\mathbf{H}_l^k)^{\text{T}} \mathbf{1}$;
            \State $\widetilde{\mathbf{B}}_l^{k} = \sqrt{\frac{\rho_l}{\alpha_l}} \mathbf{B}_l^{k}$;
            \State $\widetilde{\mathbf{JH}}_l^k = \sqrt{\frac{\rho_l}{\alpha_l}} \mathbf{JH}_{l}^{k}$;
        \EndFor
        \State $(\bm{\mu}^{k+1})^{\text{T}} = [(\bm{\mu}_1^{k+1})^{\text{T}} \ldots (\bm{\mu}_l^{k+1})^{\text{T}} \ldots (\bm{\mu}_L^{k+1})^{\text{T}}]$;
        \State $(\widetilde{\mathbf{B}}^k)^{\text{T}} = [(\widetilde{\mathbf{B}}_1^k)^{\text{T}}  \ldots (\widetilde{\mathbf{B}}_l^k)^{\text{T}} \ldots (\widetilde{\mathbf{B}}_L^k)^{\text{T}}]$;
        \State $\widetilde{\mathbf{JH}}^k = [\widetilde{\mathbf{JH}}_1^k \ldots \widetilde{\mathbf{JH}}_l^k \ldots \widetilde{\mathbf{JH}}_L^k]$;
        \State $\mathbf{UDV}^{\text{T}} = \widetilde{\mathbf{JH}}^{k} \widetilde{\mathbf{B}}^k$;
        \State $\mathbf{A}^{k+1} = \mathbf{UV}^{\text{T}}$;
        \For{$l=1\ldots L$}
            \For{$r=1\ldots R$}
                \State $\sigma_{lr}^k = ||\mathbf{b}_{l,r}^k||_2$;
                \State $\omega_{lr}^k \in \partial g(\sigma_{lr}^k)$;
                \State $\tilde{\lambda}_{lr} = \lambda_l \sqrt{J_l} \omega_{lr}^k \alpha_l/\rho_l$;
                \State $\mathbf{b}_{l,r}^{k+1} = \max(0, 1- \frac{\tilde{\lambda}_{lr}}{||(\mathbf{JH}_{l}^{k})^{\text{T}} \mathbf{a}_r^{k+1}||_2}) (\mathbf{JH}_{l}^{k})^{\text{T}} \mathbf{a}_r^{k+1}$;
             \EndFor
             \State $\mathbf{B}_{l}^{k+1} = [\mathbf{b}_{l,1}^{k+1} \ldots \mathbf{b}_{l,r}^{k+1} \ldots \mathbf{b}_{l,R}^{k+1}]$;
        \EndFor
        \State $(\mathbf{B}^{k+1})^{\text{T}} = [(\mathbf{B}_{1}^{k+1})^{\text{T}} \ldots (\mathbf{B}_{l}^{k+1})^{\text{T}} \ldots (\mathbf{B}_{L}^{k+1})^{\text{T}}]$
        \State $k=k+1$;
    \EndWhile
    \State Compute variation explained ratios.
  \end{algorithmic}
\end{algorithm}

\subsection{Variation explained ratio of the P-ESCA model}
For the quantitative data set $\mathbf{X}_l$, the parameters are $\bm{\mu}_l$, $\mathbf{A}$ and $\mathbf{B}_l$. The total variation explained ratio of the model for $\mathbf{X}_l$ is defined as $\text{varExp}_{l} = 1 - ||\mathbf{W}_l \odot (\mathbf{X}_l - \mathbf{1} \bm{\mu}_l^{\text{T}} - \mathbf{AB}_l^{\text{T}})||_F^2/||\mathbf{W}_l \odot (\mathbf{X}_l - \mathbf{1}\bm{\mu}_l^{\text{T}})||_F^2$. And the variation explained ratio for the $r^{\text{th}}$ component on $\mathbf{X}_l$ is defined as $\text{varExp}_{lr} = 1 - ||\mathbf{W}_l \odot (\mathbf{X}_l - \mathbf{1}\bm{\mu}_l^{\text{T}} - \mathbf{a}_r \mathbf{b}_{l,r}^{\text{T}})||_F^2/||\mathbf{W}_l \odot (\mathbf{X}_l - \mathbf{1}\bm{\mu}_l^{\text{T}})||_F^2$. For the binary data set, we use a similar strategy as the MOFA model \cite{argelaguet2018multi}, where the $\mathbf{H}_{l}^k$ is taken as the pseudo $\mathbf{X}_l$ during the $k^{\text{th}}$ iteration, and $\mathbf{H}_{l}^k$ rather than $\mathbf{X}_l$ is used to compute the variation explained ratios. The multiple data sets can also be taken as a single full data set. In that case the $\{1/\sqrt{\alpha_l} \}_{1}^{L}$ values are taken as the weights for them, and then we can compute the variation explained ratios of each component for this full data set. The full single data set $\widetilde{\mathbf{X}}$ and the weighting matrix $\widetilde{\mathbf{W}}$ are the column concatenation of $\{(1/\sqrt{\alpha_l}) \mathbf{X}_l \}_{1}^{L}$ and $\{ \mathbf{W}_l \}_{1}^{L}$, in which $\mathbf{X}_l$ is replaced by $\mathbf{H}^{k}_l$ if the $l^{\text{th}}$ data set is not quantitative. The offset term $\widetilde{\bm{\mu}}$ and the loading matrix $\widetilde{\mathbf{B}}$ are the row concatenation of $\{ (1/\sqrt{\alpha_l}) \bm{\mu}_l \}_{1}^{L}$ and $\{ (1/\sqrt{\alpha_l}) \mathbf{B}_l \}_{1}^{L}$ and the score matrix $\widetilde{\mathbf{A}} = \mathbf{A}$.\\

\section{Simulation process}
To evaluate the proposed model and the model selection procedure, three data sets of different data types with underlying global, local common and distinct structures are simulated. The following simulations and experiments focus on the quantitative and binary data types. We will first show the simulation of the structure $\mathbf{A}\mathbf{B}^{\text{T}}$, in which $\mathbf{B}$ is the row concatenation of $\left\{\mathbf{B}_l\right\}_{l=1}^3$, $\mathbf{B}^{\text{T}} = [\mathbf{B}_1^{\text{T}} \quad \mathbf{B}_2^{\text{T}} \quad \mathbf{B}_3^{\text{T}}]$. The structure $\mathbf{A}\mathbf{B}^{\text{T}}$ can be expressed in the SVD type as $\mathbf{A}\mathbf{B}^{\text{T}} = \mathbf{U}\mathbf{D}\mathbf{V}^{\text{T}}$ ($\mathbf{A}=\mathbf{U}$, $\mathbf{B}=\mathbf{V}\mathbf{D}$), in which $\mathbf{U}^{\text{T}}\mathbf{U} = \mathbf{I}$, $\mathbf{D}$ is a diagonal matrix, and the structured sparse pattern exists in the matrix $\mathbf{V}$. First, all the elements in $\mathbf{U}$ and $\mathbf{V}$ are simulated from the standard normal distribution. To make sure that $\mathbf{1}^{\text{T}}\mathbf{U} = \mathbf{0}$, simulated $\mathbf{U}$ is first column centered, and then it is orthogonalized by the SVD algorithm to have $\mathbf{U}^{\text{T}}\mathbf{U} = \mathbf{I}$. Also, $\mathbf{V}$ is orthogonalized by the QR algorithm to obtain $\mathbf{V}^{\text{T}}\mathbf{V} = \mathbf{I}$. In this example 21 components are predefined, 7 groups of global, local common and distinctive nature, 3 components each. The structure of these components are set in $\mathbf{V}$ as indicated below,
\begin{equation*}
\begin{aligned}
   \mathbf{V}= \left(
                 \begin{array}{c}
                   \mathbf{V}_1 \\
                   \mathbf{V}_2 \\
                   \mathbf{V}_3 \\
                 \end{array}
               \right)
               = \left(
                   \begin{array}{cccccccc}
                     \mathbf{V}_{1,1:3} & \mathbf{V}_{1,4:6} & \mathbf{V}_{1,7:9} & \mathbf{0}       & \mathbf{V}_{1,13:15} & \mathbf{0}       & \mathbf{0}      \\
                     \mathbf{V}_{2,1:3} & \mathbf{V}_{2,4:6} & \mathbf{0}       & \mathbf{V}_{2,10:12} & \mathbf{0}       & \mathbf{V}_{2,16:18} & \mathbf{0}       \\
                     \mathbf{V}_{3,1:3} & \mathbf{0}       & \mathbf{V}_{3,7:9} & \mathbf{V}_{3,10:12} & \mathbf{0}       & \mathbf{0}       & \mathbf{V}_{3,19:21} \\
                   \end{array}
                 \right),
\end{aligned}
\end{equation*}
in which $\mathbf{V}_{1,1:3}$ indicates the loadings for the first three components for data set 1, etc. After that, 21 values are sampled from $\text{N}(1,0.5)$, and their absolute values are taken as the diagonal elements of $\mathbf{D}$. Furthermore, an extra diagonal matrix $\mathbf{C}$, which has the same size as matrix $\mathbf{D}$, is used to adjust the signal to noise ratios (SNRs) in simulating different global, local common and distinct structures. Then we have $\mathbf{A}\mathbf{B}^{\text{T}} = \mathbf{U}(\mathbf{C}\odot \mathbf{D})\mathbf{V}^{\text{T}}$. In order to define the SNR, we have to specify the noise term $\mathbf{E}_l$ for the $l^{\text{th}}$ data set $\mathbf{X}_l$. If $\mathbf{X}_l$ is quantitative, all the elements in $\mathbf{E}_l$ can be sampled from $N(0,\alpha_l)$. If $\mathbf{X}_l$ is binary, according to the latent variable interpretation of logistic PCA \cite{davenport20141}, we assume there is a continuous latent matrix $\mathbf{X}_{l}^{\ast}$ underlying the binary observation $\mathbf{X}_l$, and the elements of the noise term $\mathbf{E}_l$ follow the standard logistic distribution. After the specification of the noise terms, we can adjust the diagonal elements in $\mathbf{C}$ to satisfy the predefined SNRs in simulating the global, local common and distinct structures. We restrict the diagonal elements of $\mathbf{C}$ for the same structure to share a single value to have a unique solution. For example, for the global structure $\text{C}123 = \mathbf{U}_{:,1:3}(\mathbf{C}_{1:3,1:3} \odot \mathbf{D}_{1:3,1:3}) \mathbf{V}_{:,1:3}^{\text{T}}$, the corresponding noise term is $\mathbf{E}_{123} = [\mathbf{E}_{1} \quad \mathbf{E}_{2} \quad \mathbf{E}_{3}]$, and the SNR of the global structure as defined as $\text{SNR} = \frac{||\text{C}123||_F^2}{||\mathbf{E}_{123}||_F^2}$. The SNRs for the simulation of the local common (C12, C13, C23) and distinct (D1, D2, D3) structures are defined in the same way.\\

If $\mathbf{X}_l$ is quantitative, we simply sample all the elements in $\bm{\mu}_l$ from the standard normal distribution. If $\mathbf{X}_l$ is binary, the column offset $\bm{\mu}_l$ represents the logit transformation of the marginal probabilities of binary variables. In our simulation, we will first sample $J_l$ marginal probabilities from a Beta distribution. The Beta distribution can be specified in the following way. For example, if we have 100 samples of a binary variable and we assume the marginal probability to be 0.1, this means we only observe $100 \times 0.1 = 10$ ``1''s. If we model them as Binomial observations with parameter $\pi$, and use a uniform prior distribution for $\pi$, then the posterior distribution of $\pi$ is $\pi \sim \text{Beta}(11, 91)$ \cite{gelman2013bayesian}. After generating $J_l$ marginal probabilities from this Beta distribution, the logit transformation of this vector of probabilities are set as $\bm{\mu}_l$. If $\mathbf{X}_l$ is quantitative, $\mathbf{X}_l$ is simulated as $\mathbf{X}_l = \mathbf{1}\bm{\mu}_l^{\text{T}} + \mathbf{A}\mathbf{B}_l^{\text{T}} + \mathbf{E}_l$, and all the elements of $\mathbf{E}_l$ are sampled from $N(0,\alpha_l)$. If $\mathbf{X}_l$ is binary, we have $\mathbf{\Theta}_l = \mathbf{1}\bm{\mu}_l^{\text{T}} + \mathbf{A}\mathbf{B}_l^{\text{T}}$, and all the elements of $\mathbf{X}_l$ are sampled from the Bernoulli distributions, whose probabilities are the corresponding elements in the inverse logit transformation of $\mathbf{\Theta}_l$. An equivalent way to simulate the binary $\mathbf{X}_l$ is to first generate $\mathbf{X}_l^{\ast} = \mathbf{1}\bm{\mu}_l^{\text{T}} + \mathbf{A}\mathbf{B}_l^{\text{T}} + \mathbf{E}_l$, in which all the elements in $\mathbf{E}_l$ are sampled from the standard logistic distribution. Then, all the elements in $\mathbf{X}_l$ are the binary observations of the corresponding elements of $\mathbf{X}_l^{\ast}$, $x_{ij}^{l}=1$ if $(x_{ij}^{\ast})^{l} > 0$, and $x_{ij}^{l}=0$ \textit{vise versa}. In the following sections, we will use Gaussian-Gaussian-Gaussian (G-G-G) to represent the simulation of three quantitative data sets; Bernoulli-Bernoulli-Bernoulli (B-B-B) for the simulation of three binary data sets; G-B-B for a quantitative data set and two binary data sets and G-G-B for two quantitative data sets and a binary data set.\\

\section{Evaluation matrices and model selection}
To evaluate the accuracy of the model in estimating the simulated parameters, such as $\mathbf{\Theta}_l$ and $\bm{\mu}_l$, the relative mean squared error (RMSE) is used. If, for example, the simulated parameter is $\mathbf{\Theta}$, $\mathbf{\Theta} = [\mathbf{\Theta}_{1} \quad \mathbf{\Theta}_{2} \quad \mathbf{\Theta}_{3}]$, and its estimation is $\hat{\mathbf{\Theta}}$, the RMSE is defined as $\text{RMSE}(\mathbf{\Theta}) = \frac{||\mathbf{\Theta}-\hat{\mathbf{\Theta}}||_F^2}{||\mathbf{\Theta}||_F^2}$. All of the following evaluation matrices $\text{RMSE}(\mathbf{\Theta}_l)$, $\text{RMSE}(\mathbf{\Theta})$ and $\text{RMSE}(\bm{\mu})$ will be used in the experimental section. To evaluate the recovered subspaces with respect to the simulated global common, local common and distinct structures, the modified RV coefficient \cite{smilde2008matrix} is used. If the simulated global structure is $\text{C}123$, and its estimation is $\widehat{\text{C}123}$, the similarity between the subspaces of $\text{C}123$ and $\widehat{\text{C}123}$ is calculated by the modified RV coefficient.\\

For the real data sets, we can use the cross validation (CV) error as the proxy of the prediction error to estimate the performance of the model. From each data set $\mathbf{X}_l$, we will randomly select 10\% non-missing elements as $\mathbf{X}_l^{\text{test}}$, and these selected elements in $\mathbf{X}_l$ are set to missing values. The remaining elements form the training set $\mathbf{X}_l^{\text{train}}$. For binary data, the selection of the test set samples is performed in a stratified manner to tackle the situation of unbalanced binary data. Here the test set consist of 10\% ``1''s and ``0''s which are randomly selected from $\mathbf{X}_l$ as $\mathbf{X}_l^{\text{test}}$. A P-ESCA model is constructed on the training sets $\left\{\mathbf{X}_l^{\text{train}} \right\}_{l=1}^{L}$, to obtain an estimation of $\{\hat{\mathbf{\Theta}}_l \}_1^{L}$, in which $\hat{\mathbf{\Theta}}_l = \mathbf{1}\hat{\bm{\mu}_l}^{\text{T}} + \hat{\mathbf{A}}\hat{\mathbf{B}}_l^{\text{T}}$. Then the parameters $\{\hat{\mathbf{\Theta}}_l^{\text{test}} \}_1^{L}$ corresponding to $\{ \hat{\mathbf{X}^{\text{test}}}_l \}_1^{L}$ are indexed out. The CV error for $\mathbf{X}_l$ is obtained as the negative log likelihood of using $\hat{\mathbf{\Theta}}_l^{\text{test}}$ to predict $\mathbf{X}_l^{\text{test}}$.\\

If the data sets $\left\{ \mathbf{X}_l \right\}_{l=1}^{L}$ are of the same data type, a single tuning parameter $\lambda$ is used to replace the $\left\{ \lambda_l \right\}_{l=1}^{L}$ during the model selection. First, $\left\{ \mathbf{X}_l \right\}_{l=1}^{L}$ are split into $\left\{ \mathbf{X}_l^{\text{train}} \right\}_{l=1}^{L}$ and $\left\{ \mathbf{X}_l^{\text{test}} \right\}_{l=1}^{L}$ in the same way as described above. Then $N$ $\lambda$ values are selected (with equal distance in log-space) and for each $\lambda$ value a P-ESCA model is constructed on the training sets $\left\{ \mathbf{X}_l^{\text{train}} \right\}_{l=1}^{L}$. A warm start strategy is used, in which the outputs of a previous model are used to initialize the next model with a slightly higher regularization strength. The warm start strategy has a special meaning in the current context. If some component loadings are shrunk to 0 in the previous model, they will also be 0 in the next models with higher $\lambda$ values. Thus, the search space of the next model will be constrained based on the learned structured sparse pattern in the previous model. In this way, with increasing $\lambda$, components are removed adaptively. We prefer to select the model with the minimum CV error on $\left\{\mathbf{X}_l^{\text{test}} \right\}_{l=1}^{L}$ and the corresponding value of $\lambda$ is $\lambda_{\text{opt}}$. After that we re-fit a P-ESCA model with $\lambda_{\text{opt}}$ on the full data sets $\left\{ \mathbf{X}_l \right\}_{l=1}^{L}$ and the outputs derived from the selected model with minimum CV error are used for initialization in order to preserve the learned structured sparse pattern.\\

If the data sets are of mixed data types, we prefer to use distinct tuning parameters for each data type. Suppose we have three data sets $\left\{\mathbf{X}_l \right\}_{l=1}^3$, of which $\mathbf{X}_1$ is quantitative and  $\left\{\mathbf{X}_l \right\}_{l=2}^3$ are binary. We specify two tuning parameters $\lambda_{g}$ and $\lambda_{b}$ for the loading matrices corresponding to the quantitative and binary data sets. A heuristic model selection approach, which has the same computational complexity as tuning a single parameter, can be used for the model selection. The splitting of $\left\{ \mathbf{X}_l \right\}_{l=1}^{L}$ into the training and test sets is the same as discussed above. Then again, $N$ values of $\lambda_{g}$ and $\lambda_{b}$ are selected with equal distance in log-space. For the first model, we fix $\lambda_{g}$ to be 0 or a very small value, and tune $\lambda_{b}$ in the same way as above. The model with the minimum CV error on the binary test sets $\left\{\mathbf{X}_l^{\text{test}}\right\}_{l=2}^3$ is selected, and the corresponding value of $\lambda_b$ is $\lambda_{\text{opt}}^b$. After that, $\lambda_{b}$ is fixed to $\lambda_{\text{opt}}^b$, and the outputs of the above selected model are set as the initialization for the models in the model selection of $\lambda_{g}$, which is done in the same way as described above. The model with the minimum CV error on the quantitative test set $\mathbf{X}_1^{\text{test}}$ is selected, and the corresponding value of $\lambda_g$ is $\lambda_{\text{opt}}^g$. After the model selection, we re-fit the P-ESCA model on the full data sets $\left\{\mathbf{X}_l \right\}_{l=1}^3$ with the $\lambda_{\text{opt}}^g$ and $\lambda_{\text{opt}}^b$ and again the outputs of the selected model in the model selection process are used for initialization.\\

\section{Experiments}
\subsection{Evaluating the $\alpha$ estimation procedure}
The dispersion parameters of the Bernoulli and Poisson distributions can always set to $1$, while for the Binomial distribution with $n$ experiments, it can always be set to $n$. However, for a Gaussian distribution, the dispersion parameter $\alpha$ represents the variance of the noise term, and is assumed to be known. Suppose we have a data set $\mathbf{X}_l$, we prefer to use a PCA model to estimate the $\alpha_l$ before constructing a P-ESCA model. The rank of the PCA model is selected by a missing value based cross validation procedure similar as described above. Details of the $\alpha$ estimation procedure are shown in the supplementary. After obtaining an estimation of $\hat{\alpha_l}$, it can be casted into the model or the data set can be scaled by $\sqrt{\hat{\alpha_l}}$, which is the estimated standard deviation. We simulated G-G-G, G-G-B and G-B-B data sets to test the $\alpha$ estimation procedure. The parameters in the simulation are set as $I = 100$, $J_1 = 5000$, $J_2 = 500$, $J_3 = 50$; the SNRs of the global, local common and distinct structures are all set to 1; the marginal probability is set to $0.1$ to simulate unbalanced binary data sets. The $\alpha$ estimation procedure was repeated 3 times and the average is taken as the estimation. As shown in Tab.~S2, the mean estimated dispersion parameters in different situations are quite accurate, and the estimations derived from the 3 times repetitions are very stable. \\

\subsection{An example of CV error based model selection}
We use the simulated G-G-G data sets as an example to show how the model selection is performed when multiple data sets are of the same data type. The following parameters are used in the simulation, $I = 100$, $J_1 = 1000$, $J_2 = 500$, $J_3 = 100$; the SNRs of global, local common and distinct structures are all set to 1; all the dispersion parameters $\{\alpha_l\}_1^3$ are set to be 1. The signals, which are taken as the singular values of the simulated structures, and the noises, which are taken as the singular values of the corresponding residual terms, are characterized in Fig.~S4. The true variation explained ratios of each component in every data set is computed using the simulated parameters, and is visualized in Fig.~S5. For the model selection procedure, the maximum number of iterations is set to 500; the stopping criteria is set to $\epsilon_{f} = 10^{-6}$; 30 $\lambda$ values are selected from the interval $[1,500]$ equidistant in log-space; 50 components are used in the initialization. The values of $\{ \alpha_{l} \}_{1}^{L}$ in the P-ESCA model are set to the estimated values from the above $\alpha$ estimation procedure.\\

Fig.~1 shows how the CV errors, RMSEs and the RV coefficients change with respect to $\lambda$ when a P-ESCA model with a group GDP ($\gamma=1$) penalty is used. The top figures in Fig.~1 show that the CV errors change in a similar way as the RMSEs. The model with minimum CV error has low RMSEs in estimating the simulated parameters (Fig.~1 top right) and correctly identifies the dimensions of the subspaces for the global, local common and distinct structures (Fig.~1 bottom). However, when the group lasso penalty is used this was not the case. Fig.~S6 shows that when a group lasso penalty is used, the models with minimal CV error do not coincide with the correct dimensions of the subspaces. In the model with minimum CV error, almost all the components are assigned to the global structure. This result relates to the fact that the lasso type penalty over-shrinks the non-zero parameters, and then CV error based model selection procedure tends to select a too complex model to compensate the biased parameter estimation. On the other hand, as the GDP penalty achieves nearly unbiased parameter estimation, the CV error based model selection procedure correctly identifies the correct model.\\

After the model selection, a high precision P-ESCA model ($\epsilon_{f} = 10^{-8}$) with a group GDP penalty is re-fitted on the full data sets with the value of $\lambda$ corresponding to the minimum CV error and the selected structured sparse pattern. For this selected model, the RMSEs in estimating $\mathbf{\Theta}$, $\mathbf{\Theta}_1$, $\mathbf{\Theta}_2$, $\mathbf{\Theta}_2$ and $\bm{\mu}$ are 0.0259, 0.0239, 0.0285, 0.0335 and 0.0096 respectively. The RV coefficients in estimating the global common structure $\text{C123}$ is 0.9985; local common structures $\text{C12}$, $\text{C13}$ and $\text{C23}$, 0.9977, 0.9969, 0.9953; the distinct structures $\text{D1}$, $\text{D2}$ and $\text{D3}$, 0.9961, 0.9937, 0.9779. The variation explained ratios of each component on the three data sets computed using the estimated parameters, visualized in Fig.~2, are very similar to the true ones in Fig.~S5. These values are very useful in exploring the constructed model.\\

\begin{figure}[h]
    \centering
    \includegraphics[width=\textwidth]{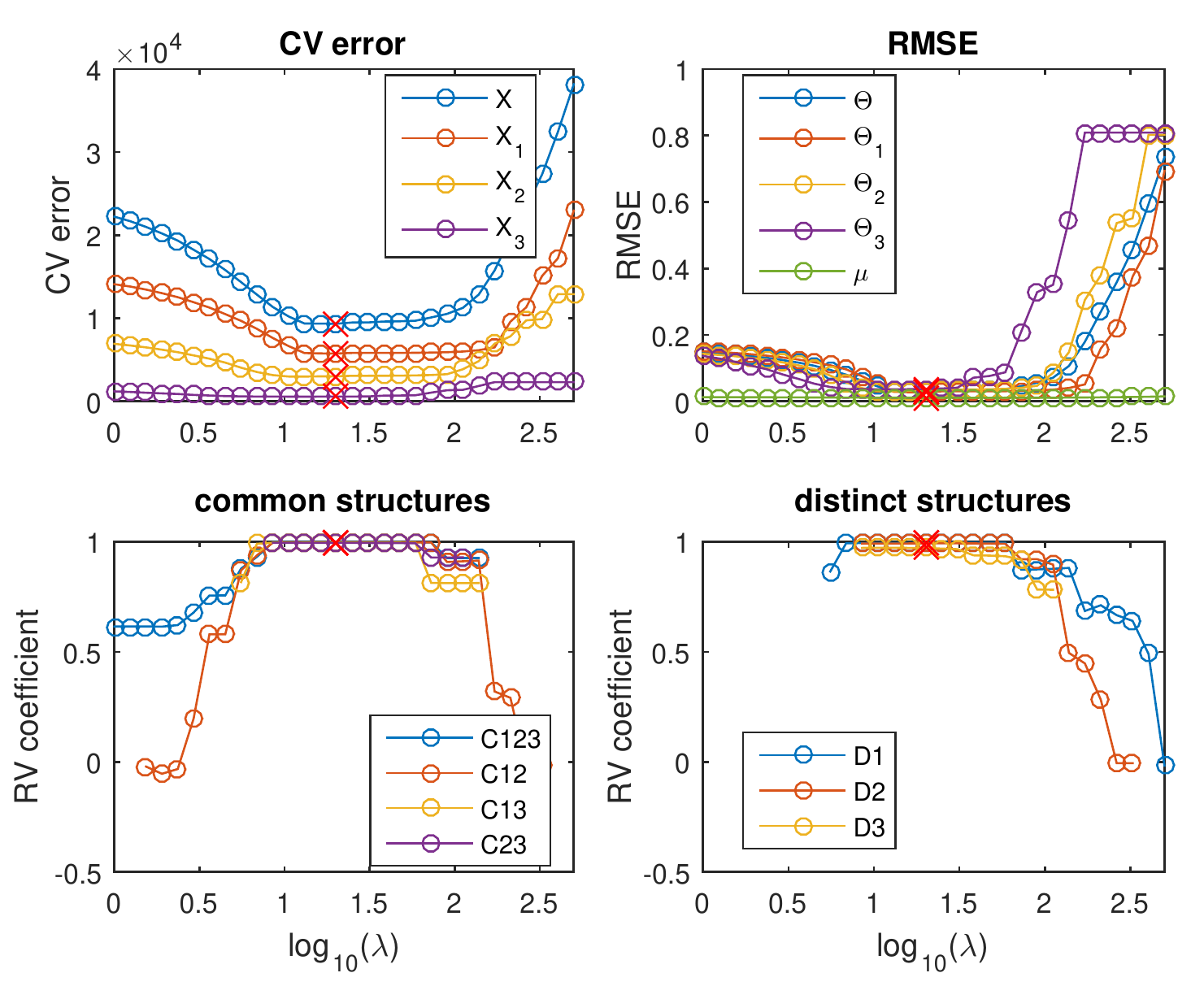}
    \caption*{\textbf{Fig.~1}: The CV errors (top left), RMSEs (top right), RV coefficients of the common structures (bottom left), and of distinct structures (bottom right) for varying $\lambda$ values for the P-ESCA model with a group GDP ($\gamma = 1$) penalty. The red cross marker indicates the model with minimum CV error.}
	\label{Fig:1}
\end{figure}

\begin{figure}[h]
    \centering
    \includegraphics[width=\textwidth]{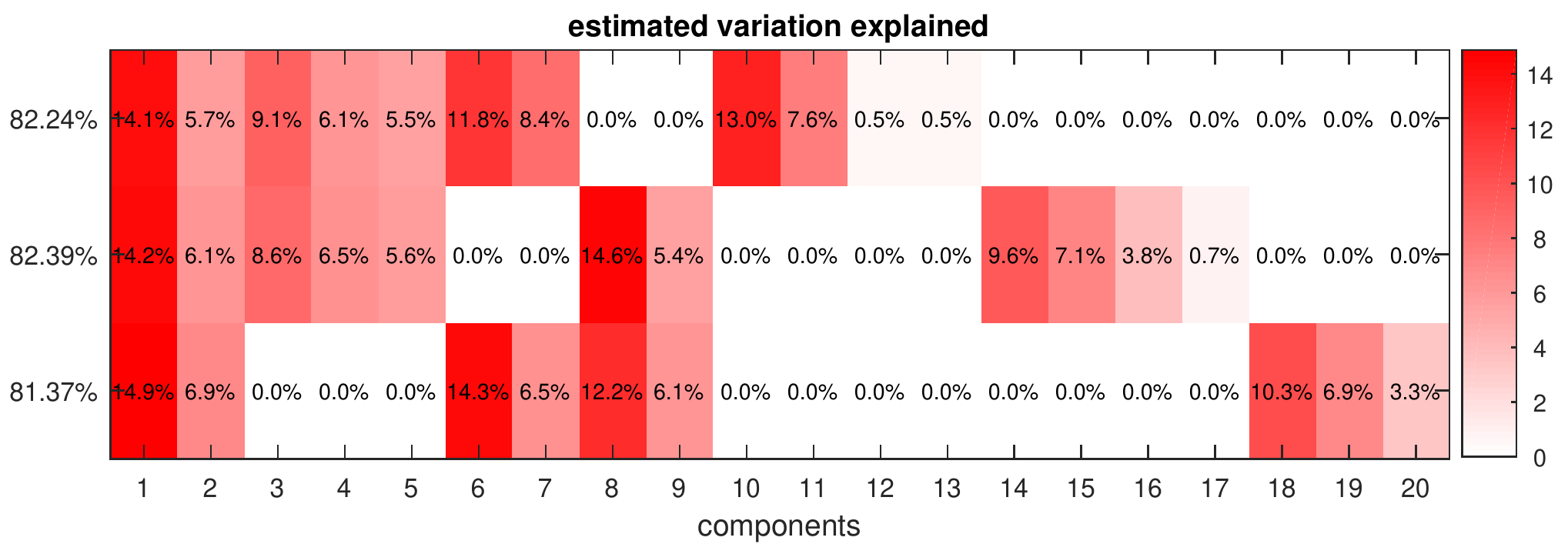}
    \caption*{\textbf{Fig.~2}: Variation explained ratios computed using the estimated parameters from the selected P-ESCA model with a group GDP penalty. From the top to the bottom, we have data sets $\mathbf{X}_1$, $\mathbf{X}_2$ and $\mathbf{X}_3$; from the left to the right, we have 20 components corresponding to the global, local common and distinct structures. The total variation explained ratios for each data set are shown on the left side of the plot, while he variation explained ratio for each component is shown inside the plot.}
	\label{Fig:2}
\end{figure}

\subsection{Full characterization of the P-ESCA model when applied to multiple quantitative data sets}
When applied to multiple quantitative data sets, our model is similar as the SLIDE model, except that we use different penalties and a different model selection procedure. The details of the differences between the two approaches are summarized in the supplementary. Since the concave GDP penalty is capable to achieve a nearly unbiased estimation of the parameters, the P-ESCA model with a group GDP penalty is expected to achieve similar performance to the two stages procedure used in the SLIDE model. Therefore, we simulated seven realistic cases by adjusting the SNRs of the simulated structures to compare the performance of these two models and their model selection procedures. The SNRs of the simulated structures corresponding to these seven cases are listed in Tab.~S3. Case 1: only the local common structures exist and they have unequal SNRs; case 2: the JIVE case, only the global common and distinct structures exist, and they are all of low SNRs; case3: all the simulated structures are of low SNRs; case 4: global common structure dominate the simulation; case 5: local common structures dominate the simulation; case 6: distinct structures dominate the simulation; case 7: none of the global, local common and distinct structures exist.\\

The following parameters are used in the G-G-G data simulations, $I = 100$, $J_1 = 1000$, $J_2 = 500$, $J_3 = 100$, all of the $\{ \alpha_l \}_1^3$ are set to 1. In order to have exactly 3 components for all the simulated structures, we reject the simulations of which the singular values of the three components of any specific structure are not 2 times larger than the singular value of the corresponding residual term. The P-ESCA model with a group GDP ($\gamma=1$) penalty is selected and re-fitted on the full data sets in the same way as above. For the SLIDE model, the simulated data sets $\{ \mathbf{X}_l \}_1^{3} $ are column centered and block-scaled by the Frobenius norm of each data set. Then the SLIDE model is selected and fitted using the default parameters. The deflated column offset term is taken as the estimated $\hat{\bm{\mu}}$. The derived loading matrices $\{ \mathbf{B}_l \}_1^3$ are re-scaled by the corresponding Frobenius norm of each data set. G-G-G data sets are simulated for all the 7 cases, and for each case, the simulation experiment (data simulation, model selection, fitting the final model) is repeated 10 times for both the P-ESCA model and the SLIDE model. The mean RV coefficients in evaluating the estimated global, local common and distinct structures and the corresponding mean estimated ranks are shown in Tab.~2, and the mean RMSEs in estimating the simulated parameters are shown in Tab.~S4. In all 7 cases, these two methods have very accurate estimation of the subspaces corresponding to the global, local common and distinct structures, and of the simulated parameters $\mathbf{\Theta}$, which is the column concatenation of $\{ \mathbf{\Theta}_l \}_1^{3} $, $\{ \mathbf{\Theta}_l \}_1^{3} $, and $\bm{\mu}$, which is row concatenation of $\{ \bm{\mu}_l \}_1^L$. For some of the cases there is a slight advantage for the P-ESCA model.\\

\begin{table}[h]
\centering
\caption*{\textbf{Tab.~2}: Mean RV coefficients and the mean rank estimates in evaluating the recovered subspaces derived from 10 experiments using the P-ESCA model and the SLIDE model for seven G-G-G simulated cases. The results are shown as in mean RV coefficient(mean rank estimation) form.}
\label{Table:2}
\begin{tabular}{lllllllll}
  \toprule
case & method & C123 & C12 & C13 & C23 & D1 & D2 & D3 \\
  \midrule
 \multirow{2}{1em}{1} & P-ESCA & 0(0)           & 0.998(3)   & 0.999(3)   & 0.999(3)   & 0(0)           & 0(0)           & 0(0)           \\
                        & SLIDE  & 0(0)           & 0.998(3)   & 0.998(3)   & 0.998(3)   & 0(0)           & 0(0)           & 0(0)           \\
 \hline
 \multirow{2}{1em}{2} & P-ESCA & 0.996(3)   & 0(0)           & 0(0)           & 0(0)           & 0.997(3)   & 0.985(3)   & 0.973(3)   \\
                        & SLIDE  & 0.996(3)   & 0(0)           & 0(0)           & 0(0)           & 0.997(3)   & 0.985(3)   & 0.973(3)   \\
 \hline
 \multirow{2}{1em}{3} & P-ESCA & 0.998(3)   & 0.997(3)   & 0.997(3)   & 0.995(3)   & 0.996(3)   & 0.994(3)   & 0.976(3)   \\
                        & SLIDE  & 0.995(3)   & 0.996(3)   & 0.993(3)   & 0.991(3)   & 0.995(3)   & 0.994(3)   & 0.976(3)   \\
 \hline
 \multirow{2}{1em}{4} & P-ESCA & 1(3)          & 1(3)          & 1(3)       & 0.999(3)   & 0.997(3)   & 0.995(3)   & 0.977(3)   \\
                        & SLIDE  & 1(3)          & 1(3)          & 0.999(3)   & 0.999(3)   & 0.997(3)   & 0.994(3)   & 0.977(3)   \\
 \hline
 \multirow{2}{1em}{5} & P-ESCA & 1(3)          & 1(3)          & 1(3)          & 1(3)          & 0.998(3)   & 0.995(3)   & 0.977(3)   \\
                        & SLIDE  & 0.999(3)      & 1(3)          & 0.999(3)   & 0.999(3)   & 0.997(3)   & 0.995(3)   & 0.977(3)   \\
 \hline
 \multirow{2}{1em}{6} & P-ESCA & 0.998(3)   & 1(3)       & 0.994(3.1) & 0.999(3)   & 0.998(2.9) & 0.999(3)   & 0.998(3)   \\
                        & SLIDE  & 0.996(3)   & 0.999(3)   & 0.998(3)   & 0.998(3)   & 0.999(3)   & 0.999(3)   & 0.997(3)   \\
 \hline
 \multirow{2}{1em}{7} & P-ESCA & 0(0)           & 0(0)           & 0(0)           & 0(0)           & 0(0)           & 0(0)           & 0(0)           \\
                        & SLIDE  & 0(0)           & 0(0)           & 0(0)           & 0(0)           & 0(0)           & 0(0)           & 0(1.4) \\
  \bottomrule
\end{tabular}
\end{table}

\subsection{Full characterization of the P-ESCA model when applied to multiple binary data sets}
The performance of the proposed P-ESCA model are fully characterized with respect to multiple binary data sets. Here we make a comparison to the MOFA model, which is the Bayesian counterpart of P-ESCA. In the P-ESCA model, the structured sparse pattern is induced through a group concave penalty, and the model selection is done through missing value based cross validation, while in the MOFA model, the structured sparse pattern is induced through the automatic relevance determination approach and the model is selected through maximizing the marginal likelihood. In addition, MOFA model also shrinks a component to be 0 when its variation explained ratios for all the data sets are less than a threshold, whose default value is 0. The details of the differences are summarized in the supplementary. For the model selection of the P-ESCA model, the range of $\lambda$ values is $[1,100]$, and the other parameters are the same as before. To give an impression of the model selection process, we also characterized how the CV errors, RMSEs and the RV coefficients change with respect to $\lambda$ in the P-ESCA model with a group GDP penalty on the simulated B-B-B data sets in Fig.~S7. For the MOFA model, the default parameters are used, but as exact sparsity cannot be achieved by the automatic relevance determination procedure used in the MOFA model, we take a component for a single data set to be 0 when the variation explained ratio of this component on this data set is less than $0.1\%$.\\

In the seven B-B-B simulations cases, we set $I=200$, and the marginal probability to be $0.1$ to simulate very unbalanced binary data sets. Other parameters are the same as in the G-G-G simulation cases. The mean RV coefficients in evaluating the estimated global, local common and distinct structures and the corresponding mean estimated ranks are shown in Tab.~3, and the mean RMSEs in estimating the simulated parameters are shown in Tab.~S5. Compared to the results derived from the P-ESCA model on the G-G-G data sets (Tab.~2), the recovered subspaces related to the global, local common and distinct structures from P-ESCA model on B-B-B data sets are less accurate with respect to RV coefficient and rank estimation, especially when the SNR of a specific structure is much lower than others (in case 4, 5, 6). However, given the fact that all the three data sets only have binary observations, the recovered subspaces are accurate enough. Furthermore, it is interesting to find that such low RMSEs in estimating $\{ \mathbf{\Theta}_l \}_1^{3} $, $\bm{\mu}$ (Tab.~S5) can be achieved solely from a model on multiple binary data sets. Although these results are a little bit counter intuitive, it is coordinate with the previous research \cite{davenport20141,song2018generalized}. According to our previous research \cite{song2018generalized}, this result mainly relates to the fact that the GDP penalty can achieve nearly unbiased parameter estimation. On the other hand, the RMSEs in estimating the simulated parameters from the MOFA model (Tab.~S5) are much larger. Especially for the estimation of the simulated column offset term, all the elements in the estimated $\hat{\bm{\mu}}$ from the MOFA model are very close to 0, and are far away from the simulated $\bm{\mu}$. However, the recovered subspaces from the MOFA model are comparable to the results derived from the P-ESCA model (Tab.~3).\\

\begin{table}[h!]
\centering
\caption*{\textbf{Tab.~3}: Mean RV coefficients and mean rank estimations of recovered subspaces derived from 10 repeated simulation experiments using the P-ESCA model and the MOFA model for seven B-B-B cases. For case 7, a one component MOFA model is selected, however, results cannot be extracted when the offset term is included.}
\label{Table:3}
\begin{tabular}{lllllllll}
  \toprule
case & method & C123 & C12 & C13 & C23 & D1 & D2 & D3 \\
  \midrule
 \multirow{2}{0.5em}{1} & P-ESCA & 0(0)        & 0.993(2.9) & 0.994(2.9) & 0.991(2.5) & 0(0.2)     & 0(0.5)       & 0(0) \\
                           & MOFA   & 0(0)        & 0.834(2.1) & 0.984(3.2) & 0.989(3.2) & 0(1.4)     & 0(1  )       & 0(0)  \\
 \hline
 \multirow{2}{0.5em}{2} & P-ESCA & 0.993(2.6)  & 0(0.4)     & 0(0)       & 0(0  )     & 0.990(3  ) & 0.982(3)     & 0.914(3) \\
                           & MOFA   & 0.959(2.7)  & 0(0  )     & 0(0.1)     & 0(0.2)     & 0.964(3.2) & 0.975(3.1)   & 0.885(2.6) \\
 \hline
 \multirow{2}{0.5em}{3} & P-ESCA & 0.956(1.9)  & 0.959(4) & 0.972(1.8) & 0.939(1.9) & 0.967(4.3) & 0.945(4.1)   & 0.878(2.6) \\
                           & MOFA   & 0.940(2.6)  & 0.925(2.3) & 0.977(3.2)   & 0.956(3.1) & 0.936(3.6) & 0.934(3.7)   & 0.848(2.3) \\
 \hline
 \multirow{2}{0.5em}{4} & P-ESCA & 0.992(2.3)  & 0.988(3.3) & 0.981(2.4) & 0.980(2.2) & 0.831(3.6) & 0.838(3.2 )   & 0.151(0.2) \\
                           & MOFA   & 0.986(2.9)  & 0.955(2.9) & 0.990(3  ) & 0.985(2.9) & 0.960(2.6) & 0.929(2.3)   & 0.220(0.3)      \\
 \hline
 \multirow{2}{0.5em}{5} & P-ESCA & 0.980(2.1)  & 0.990(3.8) & 0.991(2.5)     & 0.986(2.6)   & 0.916(3.4) & 0.808(2.3)   & 0.074(0.1)\\
                           & MOFA   & 0.915(3.1)  & 0.956(2.8) & 0.991(2.9)     & 0.984(3  )   & 0.878(2.6)   & 0.917(2  )   & 0.193(0.3)      \\
 \hline
 \multirow{2}{0.5em}{6} & P-ESCA & 0.192(0.2)  & 0.981(4.7) & 0.984(2.3) & 0.979(2.6) & 0.991(4.6) & 0.988(3   )   & 0.963(2.8) \\
                           & MOFA   & 0.525(1.1)  & 0.949(2.1) & 0.980(3.7) & 0.979(3.4)  & 0.978(4.3) & 0.977(4.2)   & 0.953(3.1) \\
 \hline
 \multirow{2}{0.5em}{7} & P-ESCA & 0(0)        & 0(0)        & 0(0)        & 0(0)        & 0(0)        & 0(0)        & 0(0) \\
                           & MOFA  & NA          & NA           & NA           & NA         & NA         &NA         & NA \\
  \bottomrule
\end{tabular}
\end{table}

\subsection{Full characterization of the P-ESCA model when applied to multiple data sets of mixed data types}
The proposed P-ESCA model is also fully characterized on the simulated multiple data sets of mixed quantitative and binary data types. Both G-B-B and G-G-B data sets are simulated for all the seven simulation cases. We set $I=200$, all of $\{ \alpha_l \}_1^3$ to be 1, the marginal probability in simulating unbalanced data sets to be $0.1$. Other parameters are the same as above. The range of $\lambda$ values for loadings related to the quantitative data sets is $[1,500]$, and for loadings related to binary data sets is $[1,100]$. The mean RV coefficients of the estimated global, local common and distinct structures and the corresponding mean ranks estimation from the P-ESCA and the MOFA model in the seven G-B-B simulation cases are shown in Tab.~4, for the G-G-B simulation the results are shown in Tab.~5. The mean RMSEs in estimating the simulated parameters are shown in Tab.~S6, for the G-B-B simulations are in Tab.~S7. Similar to the previous results of B-B-B simulations, the P-ESCA model can achieve quite accurate estimates of the subspaces related to the global, local common and distinct structures (Tab.~4, Tab.~5) when the SNRs of different structures are relative equal. However, when the SNR of a specific structure is very low compared to others (in case 4, 5, 6), the P-ESCA model has difficulty for its recovery. However, compared to the MOFA model, P-ESCA can achieve better results with respect to the recovered subspaces (Tab.~4, Tab.~5) and estimation of the simulated parameters (Tab.~S6, Tab.~S7) in G-B-B and G-G-B simulations.\\

\begin{table}[h!]
\centering
\caption*{\textbf{Tab.~4}: Mean RV coefficients and mean rank estimations of the recovered subspaces derived from simulation experiments using the P-ESCA model and the MOFA model for seven G-B-B cases.}
\label{Table:4}
\begin{tabular}{lllllllll}
  \toprule
case & method & C123 & C12 & C13 & C23 & D1 & D2 & D3 \\
  \midrule
 \multirow{2}{0.5em}{1}    & ESCA  & 0(0)        & 0.997(2.8)   & 0.987(2.3)   & 0.993(3) & 0(0.9)     & 0(0)       & 0(0) \\
                           & MOFA  & 0(0)        & 0.826(2.5)   & 0.978(3)   & 0.973(3.7) & 0(1.6)     & 0(0.5)     & 0(0)  \\
 \hline
 \multirow{2}{0.5em}{2}    & ESCA  & 0.978(2.3)  & 0(0.5)       & 0(0)       & 0(0)       & 0.993(3.2) & 0.981(3) & 0.918(2.9) \\
                           & MOFA  & 0.984(2.7)  & 0(0.1)       & 0(0)       & 0(0.2)     & 0.533(4.2) & 0.975(3) & 0.895(2.7) \\
 \hline
 \multirow{2}{0.5em}{3}    & ESCA  & 0.975(2)  & 0.972(3.9)   & 0.945(1.5)   & 0.974(2.2) & 0.932(4.6)   & 0.968(3.8)   & 0.892(2.6) \\
                           & MOFA  & 0.914(3)  & 0.879(2.7)   & 0.962(2.7)   & 0.971(3.1) & 0.475(4.6)   & 0.970(2.9)   & 0.860(2.5) \\
 \hline
 \multirow{2}{0.5em}{4}    & ESCA  & 0.998(2.7)  & 0.995(2.9)     & 0.917(1.6)     & 0.991(2.4)   & 0.547(4.8)   & 0.909(3.3)   & 0(0) \\
                           & MOFA  & 0.856(3.7)  & 0.547(2)     & 0.788(3.7)     & 0.990(3)   & 0.378(4.9)   & 0.935(2.8)   & 0.398(0.6)   \\
 \hline
 \multirow{2}{0.5em}{5}    & ESCA  & 0.982(2.1)  & 0.995(3.4)     & 0.994(2.2)     & 0.994(2.8)   & 0.698(4.3)   & 0.929(3.1) & 0.164(0.2)\\
                           & MOFA  & 0.677(3.3)  & 0.691(2)     & 0.916(3.3)     & 0.991(3.1)   & 0.316(5.2)   & 0.835(2.8) & 0.475(0.8)   \\
 \hline
 \multirow{2}{0.5em}{6}    & ESCA  & 0(0)      & 0.980(5.1)       & 0.971(1.8)     & 0.989(2.5)   & 0.989(5.1)   & 0.992(3.5)   & 0.966(2.9) \\
                           & MOFA  & 0.624(1.4)  & 0.750(1.9)     & 0.899(3.9)     & 0.985(3.4)   & 0.837(6.2)   & 0.978(4.3)   & 0.954(2.9) \\
 \hline
 \multirow{2}{0.5em}{7} & ESCA  & 0(0)        & 0(0)         & 0(0)         & 0(0)       & 0(0)       & 0(0)       & 0(0) \\
                           & MOFA  & NA          & NA           & NA           & NA         & NA         &NA         & NA    \\
  \bottomrule
\end{tabular}
\end{table}

\begin{table}[h!]
\centering
\caption*{\textbf{Tab.~5}: Mean RV coefficients and mean rank estimations of the recovered subspaces derived from 10 simulation experiments using the P-ESCA model and the MOFA model for seven G-G-B cases.}
\label{Table:5}
\begin{tabular}{lllllllll}
  \toprule
case & method & C123 & C12 & C13 & C23 & D1 & D2 & D3 \\
  \midrule
 \multirow{2}{0.5em}{1} & P-ESCA & 0(0)        & 0.998(3) & 0.997(2.4) & 0.999(2.9) & 0(0.6)     & 0(0.1)       & 0(0) \\
                           & MOFA   & 0(0)        & 0.528(3.8) & 0.998(2.8) & 0.998(2.9) & 0(0.4)     & 0(0.3)       & 0(0)  \\
 \hline
 \multirow{2}{0.5em}{2} & P-ESCA & 0.971(2.4)  & 0(0.6)     & 0(0)       & 0(0)     & 0.998(3) & 0.995(3)     & 0.920(2.9) \\
                           & MOFA   & 0.987(2.8)  & 0(1.2)     & 0(0)       & 0(0)       & 0.997(3)   & 0.994(3)     & 0.899(2.8) \\
 \hline
 \multirow{2}{0.5em}{3} & P-ESCA & 0.984(2.2)  & 0.977(3.8) & 0.979(2.2) & 0.981(2.2) & 0.970(3.8) & 0.968(3.8)   & 0.922(3) \\
                           & MOFA   & 0.977(2.7)  & 0.524(4.3) & 0.993(2.8)   & 0.989(2.9) & 0.989(3.2) & 0.980(3.1)   & 0.888(2.3) \\
 \hline
 \multirow{2}{0.5em}{4} & P-ESCA & 0.996(3)    & 0.965(3  ) & 0.997(2.6) & 0.996(2.4) & 0.941(3.4) & 0.899(3.6)   & 0.844(2.4) \\
                           & MOFA   & 0.955(4)    & 0.673(3.1) & 0.903(2.9) & 0.997(2.9) & 0.920(3) & 0.983(3.1)   & 0.703(1.2)      \\
 \hline
 \multirow{2}{0.5em}{5} & P-ESCA & 0.996(2.4)  & 0.998(3.6) & 1(2.6)     & 0.999(2.6)   & 0.978(3.4) & 0.952(3.4)   & 0.808(1.8)\\
                           & MOFA   & 0.761(3.9)  & 0.715(3.2) & 0.999(3)   & 0.998(2.9) & 0.995(3.1)   & 0.982(3.2)   & 0.494(0.7)      \\
 \hline
 \multirow{2}{0.5em}{6} & P-ESCA & 0.348(0.5)  & 0.984(5.5) & 0.992(2.4) & 0.996(2.5) & 0.997(3.6) & 0.997(3.5)   & 0.970(3) \\
                           & MOFA   & 0.894(2)  & 0.949(5) & 0.925(3.2) & 0.972(3)  & 0.933(3) & 0.973(3.1)   & 0.960(2.9) \\
 \hline
 \multirow{2}{0.5em}{7} & P-ESCA & 0(0)        & 0(0)        & 0(0)        & 0(0)        & 0(0)        & 0(0)        & 0(0) \\
                           & MOFA   & NA          & NA           & NA           & NA         & NA         &NA         & NA \\
  \bottomrule
\end{tabular}
\end{table}

\section{Real data analysis}
We applied the P-ESCA model on the chronic lymphocytic leukaemia (CLL) data set \cite{dietrich2018drug,argelaguet2018multi}, which was used in the paper of the MOFA model, to give an example of the real data analysis. For the 200 samples in the CLL data set, not all of them are fully characterized for all the measurements. Drug response data has 184 samples and 310 variables; DNA methylation data, 196 samples and 4248 variables; transcriptome data, 136 samples and 5000 variables; mutation data, 200 samples and 69 binary variables. The missing pattern of the CLL data sets is visualized in Fig.~S8. Except for the missing values related to the samples that were not measured by a specific platform, there are also some selected variables missing in the mutation data (Fig.~S8). All the quantitative data sets are first column centered and scaled by the sample standard deviation of each variable. After that, the dispersion parameters of the quantitative data sets are estimated by the $\alpha$ estimation procedure. Rank estimation of each single data set was performed three times and results are shown in Tab.~S8. The P-ESCA model with a GDP ($\gamma=1$) is selected and re-fitted on the CLL data sets in the same way as described above. The initial number of components is set to 50. The selected model has 41 components, and if we take each loading vector related to a single data set in a component as a group, there are 51 non-zero loading groups. The model selection results are shown in Fig.~S9. Since the variation explained ratios of 41 components are difficult to visualize, we only show the components (Fig.~3), whose variation explained ratio are larger than 2\% for at least one data set. The above procedure (processing, model selection, fitting the final model) is repeated 5 times to test its stability. The Pearson coefficient matrix for the 5 estimations of the $\hat{\bm{\mu}}$ and the RV coefficient matrices for the 5 estimations of the $\hat{\mathbf{A}}$, $\hat{\mathbf{B}}$ and $\hat{\mathbf{\Theta}}$ are shown in Fig.~S10.\\

In \cite{argelaguet2018multi}, a 10 components MOFA model is selected on the CLL data sets. The variation explained plots of the 10 components MOFA model, reproduced from \cite{argelaguet2018multi}, is shown in Fig.~S11. There is some overlap between the two models (Fig.~3, Fig.~S11). Both models have one strong common component in which all data sets participate, and a common component in which two (P-ESCA) or three (MOFA) data sets participate. Furthermore the drug response and the transcriptomic (mRNA) data have extra distinct components. The variation explained is somewhat higher for the P-ESCA model which also uses extra components. The amount of variation explained is the highest for the drug response and mRNA data sets. The main difference between the models is the fact that P-ESCA only finds a single component relevant for the binary mutation data while MOFA finds two. The comparison of the two models with respect to the estimated $\hat{\bm{\mu}}$ is infeasible because the column offset term is not included in this 10 components MOFA model. In general the P-ESCA result is more complex than the results in \cite{argelaguet2018multi} in terms of number of selected components and variation explained. However, this is mainly because, during the model selection of \cite{argelaguet2018multi}, the minimum variation explained threshold is set to 2\%. If we set the threshold to the default value 0\%, and set the initial number of components to be 50, and other parameters are kept the same, a 50 components MOFA model is selected.\\

\begin{figure}[h]
    \centering
    \includegraphics[width=\textwidth]{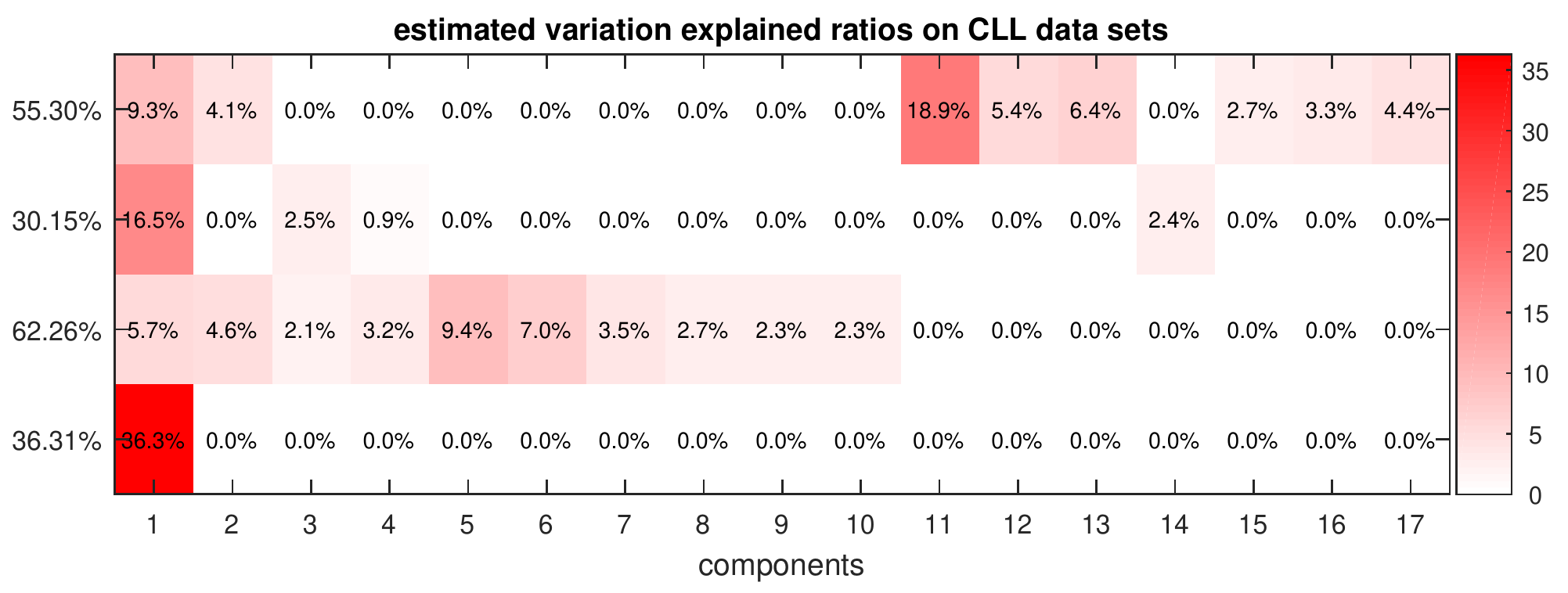}
    \caption*{\textbf{Fig.~3}: Variation explained ratios computed using the estimated parameters from the selected P-ESCA model on CLL data sets. From the top to the bottom, the data sets are drug response, methylation, transcriptome and mutation data.}
	\label{Fig:3}
\end{figure}

\section{Discussion}
In this paper, we generalized an exponential family SCA (ESCA) model for the data integration of multiple data sets of mixed data types. Then, we introduced the nearly unbiased group concave penalty to induce structured sparsity pattern on the loading matrices of the ESCA model to separate the global, local common and distinct variation. An efficient MM algorithm with analytical form updates for all the parameters was derived to fit the proposed group concave penalty penalized ESCA (P-ESCA) model. In addition, a missing value based cross validation procedure is developed for the model selection. In many different realistic simulations (different SNR levels, and combinations of quantitative and or binary data sets of different), the P-ESCA model and the model selection procedure work well with respect to recovering the subspaces related to the global, local common and distinct structures, and the estimation of the simulated parameters.\\

The performance of the P-ESCA model and the cross validation based model selection procedure relate to the fact that the used group concave penalty can achieve nearly unbiased estimation of the parameters while generating sparse solutions. The nearly unbiased parameter estimation makes the P-ESCA model have high accuracy in the estimation of the simulated parameters, and the cross validation error based model selection procedure is consistent. Another key point of the model selection procedure is that the randomly sampled $10\%$ non-missing elements are usually a typical set of elements from the population. This makes the CV error a good proxy of the prediction error of the model. The rank estimation in different repetitions of the model selection procedure is robust and only differ slightly with respect to the very weak components. \\

When applied to multiple quantitative data sets, the proposed P-ESCA model can achieve slightly better performance than the SLIDE model in recovering the subspaces of the simulated structures and in estimating the simulated parameters. Also, since missing value problems (missing values in a single data set, or missing complete samples in one or some of the data sets) are very common in practice, the option of tackling missing values is a big advantage. In the P-ESCA model and its model selection procedure, the effect of missing values is masked in a very natural way, making full use of the available data sets. When applied to the multiple binary data sets or the mixed quantitative and binary data sets, the proposed P-ESCA model has better performance than the MOFA model in recovering the subspaces of the simulated structures and in estimating the simulated parameters. Furthermore, the exact orthogonality constraint can be achieved in the P-ESCA model, which is crucial for the uniqueness of the recovered subspaces related to the global, local common and distinct variation.\\

\bibliographystyle{ieeetr}

\bibliography{reference}

\newpage
\section*{Supplementary files}

\subsection*{$\alpha$ estimation using PCA}
Before constructing an ESCA or P-ESCA model, the dispersion parameter $\alpha$ of a quantitative data set $\mathbf{X}$, which is the variance of the residual term, is assumed to be known. Assume the column centered quantitative data set is $\mathbf{X}$($I \times J$), and the PCA model of $\mathbf{X}$ can be expressed as $\mathbf{X} = \mathbf{AB}^{\text{T}} + \mathbf{E}$. $\mathbf{A}$($I \times R$) and $\mathbf{B}$($J \times R$) are the score and loading matrix respectively; $\mathbf{E}$($I \times J$) is the residual term and elements in $\mathbf{E}$, $\epsilon_{ij} \sim N(0,\alpha)$; $R$ is the true low rank of $\mathbf{X}$. In order to tackle the potential missing value problem, we also introduce the weighting matrix $\mathbf{W}$ in the same way as above. The maximum likelihood estimation of $\alpha$ can be expressed as $\hat{\alpha}_{\text{mle}} = \frac{1}{||\mathbf{W}||_0} ||\mathbf{W} \odot (\mathbf{X} - \mathbf{AB}^{\text{T}})||_{F}^2$, in which $||\mathbf{W}||_0$ is the number of non-missing elements in $\mathbf{W}$. Since this is a biased estimation of $\alpha$, we can adjust the estimation according to the degree of freedom as $\hat{\alpha} = \frac{1}{||\mathbf{W}||_0 - (I+J)R} ||\mathbf{W} \odot (\mathbf{X} - \mathbf{AB}^{\text{T}})||_{F}^2$. The parameters $R$, $\mathbf{A}$ and $\mathbf{B}$ are estimated as follows.\\

We select the rank $R$ using a similar model selection strategy as in the main text. We first split $\mathbf{X}$ into $\mathbf{X}^{\text{test}}$ and $\mathbf{X}^{\text{train}}$ in the same way as in the main text. Then, a series of PCA models with different number of components are constructed on $\mathbf{X}^{\text{train}}$, and the CV error is defined as the least square error in fitting $\mathbf{X}^{\text{test}}$. After that $\hat{R}$ is set to the number of components of the model with the minimum CV error. Then a rank $\hat{R}$ PCA model is constructed on the full data $\mathbf{X}$, and we get an estimate of $\hat{\mathbf{A}}$ and $\hat{\mathbf{B}}$. Then $\hat{\alpha}$ is set to $\hat{\alpha} = \frac{1}{||\mathbf{W}||_0 - (I+J)\hat{R}} ||\mathbf{W} \odot (\mathbf{X} - \hat{\mathbf{A}}\hat{\mathbf{B}}^{\text{T}})||_{F}^2$. The EM type algorithm used to fit the PCA model with the option of missing values is implemented in Matlab in the same way as in \cite{kiers1997weighted}.\\

\subsection*{The difference between the SLIDE model and the P-ESCA model when applied to multiple quantitative data sets}
\begin{itemize}
  \item Different processing steps. The SLIDE model does column centering and block scaling using the Frobenius norm of the corresponding data set to preprocess the data. Then the relative weights of the data sets in the SCA model are set to 1. On the other hand, we estimate the dispersion parameter (variation of the noise term) of each data set and the inverse of the estimated dispersion parameter is equivalent to the relative weight of the data sets in the SCA model.
  \item Different penalty terms. The SLIDE model uses the group lasso penalty to induce the structured sparsity. Because of the block scaling processing step, there is no weight $\left\{ \sqrt{J_l} \right\}_{l=1}^L$ on the group lasso penalty to accommodate for the potential unequal number of variables in different data sets. On the other hand, the weighted group concave penalty is used in the P-ESCA model.
  \item Option for missing values. The option of tackling the missing value problem is not included in the SLIDE model.
  \item Different model selection procedures. The SLIDE model uses a two stages approach to do model selection, while our model selection approach is as described as in the main text.
\end{itemize}

\subsection*{The difference between the MOFA model and the P-ESCA model}
\begin{itemize}
  \item Different origins. Although these two methods are similar with respect to what they can do, they have different origins. The MOFA model is developed in the Bayesian probabilistic matrix factorization framework in the same line as the group factor analysis model and the factor analysis model, while the P-ESCA model is derived in the deterministic matrix factorization framework in the same line as the SLIDE model, the SCA model and the PCA model.
  \item Different ways in inducing structured sparsity. In the P-ESCA model, the structured sparse pattern is induced through a group concave penalty, while in the MOFA model, it is induced through the automatic relevance determination approach. The group concave penalty can shrink a group of elements to be exactly 0, while the automatic relevance determination cannot achieve exact sparsity. In addition, MOFA model also shrinks a component to be 0 when its variation explained ratios for all the data sets are less than a threshold, whose default value is 0.
  \item Different model selection procedures. The P-ESCA model is selected by a missing value based CV approach; while the selection of a MOFA model relies on maximizing the marginal likelihood. In theory, maximizing the marginal likelihood has no difficulty in tuning multiple parameters, while the CV based model selection procedure is infeasible for such task.
  \item Orthogonality constraint. The orthogonality constraint $\mathbf{A}^{\text{T}}\mathbf{A} = \mathbf{I}$ can only be achieved in the P-ESCA model. Whether this property is meaningful or not depends on the specific research question. However, the constraint is crucial for the proof of the uniqueness of the recovered subspaces corresponding to the global, local common and distinct variation.
\end{itemize}

\newpage

\subsection*{Supplementary tables and figures}
\begin{table}[h!]
\centering
\caption*{\textbf{Tab.~S1}: A list of log-partition functions and their first and second order derivatives for the Gaussian, Bernoulli and Poisson
distributions. $\theta$ indicates the natural parameter.}
\label{Table:S1}
\begin{tabular}{cccc}
  \toprule
Distribution & $b(\theta)$ & $b^{'}(\theta)$ & $b^{''}(\theta)$ \\
  \midrule
Gaussian     & $\frac{\theta^2}{2}$    & $\theta$ & 1 \\
Bernoulli    & $\log(1+\exp(\theta))$ & $\frac{\exp(\theta)}{1+\exp(\theta)}$ & $\frac{\exp(\theta)}{(1+\exp(\theta))^2}$ \\
Poisson      & $\exp(\theta)$ & $\exp(\theta)$ & $\exp(\theta)$ \\
  \bottomrule
\end{tabular}
\end{table}

\begin{table}[h!]
\centering
\caption*{\textbf{Tab.~S2}: Results of the $\alpha$ estimation procedure. $1^{g}$ indicates that a Gaussian distribution is used and $\alpha_l = 1$; $b$ indicates the Bernoulli distribution.  The estimated dispersion parameter $\hat{\alpha_l}$ and the corresponding times are shown as $\text{mean} \pm \text{std} (\text{seconds})$. When the estimated ranks are the same in each of the three times CV procedure is repeated, the corresponding standard deviation is 0.}
\label{Table:S2}
\begin{tabular}{llllll}
 \hline
   $\alpha_1$ & $\alpha_2$ & $\alpha_3$ & $\hat{\alpha_1}$($\text{time}$) & $\hat{\alpha_2}$($\text{time}$) & $\hat{\alpha_3}$($\text{time}$)) \\
 \hline
  $1^{g}$   & $1^{g}$  & $1^{g}$ & 0.9920  $\pm$ 0 (9.01)  & 1.0029  $\pm$ 0 (2.46) & 1.0183 $\pm$ 0 (17.06) \\
  $100^{g}$ & $25^{g}$ & $1^{g}$ & 99.7148 $\pm$ 0.7469 (10.66) & 24.9525 $\pm$ 0 (2.96) & 1.1609 $\pm$ 0.2437 (76.33) \\
  \hline
  $1^{g}$   & $1^{g}$  & $b$     & 0.9892  $\pm$ 0 (10.92)  & 0.9793  $\pm$ 0 (2.80) &  \\
  $100^{g}$ & $25^{g}$ & $b$     & 99.8457 $\pm$ 0 (10.79)  & 24.7688 $\pm$ 0 (3.43) &  \\
  \hline
  $1^{g}$   & $b$ & $b$     & 0.9896   $\pm$ 0 (10.90) &  &  \\
  $100^{g}$ & $b$ & $b$     & 100.1774 $\pm$ 0 (10.75) &  &  \\

  \hline
\end{tabular}
\end{table}

\begin{table}[h!]
\centering
\caption*{\textbf{Tab.~S3}: Seven simulation cases used to evaluate the proposed P-ESCA model. For each simulation case, the corresponding SNRs in simulating the global structure $\text{C123}$, local common structures, $\text{C12}$, $\text{C13}$, $\text{C23}$, and distinct structures $\text{D1}$, $\text{D2}$, $\text{D3}$, are give. If the SNR of a specific structure is 0, it means this structure does not exist in the simulation.}
\label{Table:S3}
\begin{tabular}{llllllll}
  \toprule
case & $\text{C123}$ & $\text{C12}$ & $\text{C13}$ & $\text{C23}$ & $\text{D1}$ & $\text{D2}$ & $\text{D3}$ \\
  \midrule
 1   & 0  & 1  & 2  & 3   & 0  & 0  & 0   \\
 2   & 1  & 0  & 0  & 0   & 1  & 1  & 1   \\
 3   & 1  & 1  & 1  & 1   & 1  & 1  & 1   \\
 4   & 10 & 5  & 5  & 5   & 1  & 1  & 1   \\
 5   & 5  & 10 & 10 & 10  & 1  & 1  & 1   \\
 6   & 1  & 5  & 5  & 5   & 10 & 10 & 10   \\
 7   & 0  &  0 & 0  & 0   & 0  & 0  & 0   \\
  \bottomrule
\end{tabular}
\end{table}

\begin{table}[h!]
\centering
\caption*{\textbf{Tab.~S4}: Mean RMSEs in estimating the simulated parameters $\mathbf{\Theta}$, $\left\{ \mathbf{\Theta} \right\}_{l=1}^3$ and $\bm{\mu}$, derived from repeating the experiments 10 time using the P-ESCA model and the SLIDE model for seven G-G-G simulation cases.}
\label{Table:S4}
\begin{tabular}{lllllll}
  \toprule
case & method & $\text{RMSE}(\mathbf{\Theta})$ & $\text{RMSE}(\mathbf{\Theta}_1)$ & $\text{RMSE}(\mathbf{\Theta}_2)$ & $\text{RMSE}(\mathbf{\Theta}_3)$ & $\text{RMSE}(\bm{\mu})$ \\
  \midrule
 \multirow{2}{0.5em}{1}& P-ESCA  &0.0167    &0.0181    &0.0152    &0.0135    &0.0102  \\
                          & SLIDE &0.0178    &0.0194    &0.0161    &0.0145    &0.0102  \\
 \hline
 \multirow{2}{0.5em}{2}& P-ESCA  &0.0251    &0.0241    &0.0255    &0.0334    &0.0100 \\
                          & SLIDE &0.0269    &0.0259    &0.0273    &0.0349    &0.0100 \\
 \hline
 \multirow{2}{0.5em}{3}& P-ESCA  &0.0274    &0.0266    &0.0278    &0.0333    &0.0097 \\
                          & SLIDE &0.0298    &0.0290    &0.0301    &0.0366    &0.0097 \\
 \hline
 \multirow{2}{0.5em}{4}& P-ESCA  &0.0064    &0.0062    &0.0065    &0.0076    &0.0099 \\
                          & SLIDE &0.0068    &0.0066    &0.0069    &0.0081    &0.0099 \\
 \hline
 \multirow{2}{0.5em}{5}& P-ESCA  &0.0052    &0.0051    &0.0052    &0.0063    &0.0099  \\
                          & SLIDE &0.0055    &0.0054    &0.0056    &0.0067    &0.0099 \\
 \hline
 \multirow{2}{0.5em}{6}& P-ESCA  &0.0064    &0.0062    &0.0065    &0.0078    &0.0100 \\
                          & SLIDE &0.0068    &0.0066    &0.0068    &0.0082    &0.0100 \\
 \hline
 \multirow{2}{0.5em}{7}& P-ESCA  &0.0099    &0.0097    &0.0104    &0.0098    &0.0099 \\
                          & SLIDE &0.0132    &0.0097    &0.0104    &0.0658    &0.0099 \\
  \bottomrule
\end{tabular}
\end{table}

\begin{table}[h!]
\centering
\caption*{\textbf{Tab.~S5}: Mean RMSEs in estimating the simulated parameters $\mathbf{\Theta}$, $\left\{ \mathbf{\Theta} \right\}_{l=1}^3$ and $\bm{\mu}$ derived from repeating the experiments 10 times using the P-ESCA model and the MOFA model for seven B-B-B simulation cases.}
\label{Table:S5}
\begin{tabular}{lllllll}
  \toprule
case & method & $\text{RMSE}(\mathbf{\Theta})$ & $\text{RMSE}(\mathbf{\Theta}_1)$ & $\text{RMSE}(\mathbf{\Theta}_2)$ & $\text{RMSE}(\mathbf{\Theta}_3)$ & $\text{RMSE}(\bm{\mu})$ \\
  \midrule
 \multirow{2}{0.5em}{1}& P-ESCA &0.0530    &0.0450    &0.0498    &0.1218    &0.0265  \\
                          & MOFA   &0.4762    &0.5004    &0.4518    &0.4130    &0.9999  \\
 \hline
 \multirow{2}{0.5em}{2}& P-ESCA &0.0528    &0.0488    &0.0511    &0.1009    &0.0223 \\
                          & MOFA   &0.5951    &0.5911    &0.5936    &0.6432    &1.0000 \\
  \hline
 \multirow{2}{0.5em}{3}& P-ESCA   &0.0830    &0.0651    &0.0775    &0.2922    &0.0331 \\
                          & MOFA    &0.5037    &0.4965    &0.5077    &0.5558    &0.9999 \\
 \hline
 \multirow{2}{0.5em}{4}& P-ESCA &0.1080    &0.0673    &0.1240    &0.4298    &0.0731 \\
                          & MOFA   &0.3297    &0.3233    &0.3322    &0.3805    &0.9999 \\
 \hline
 \multirow{2}{0.5em}{5}& P-ESCA &0.1225    &0.0750    &0.1506    &0.4546    &0.0860  \\
                          & MOFA   &0.3267    &0.3196    &0.3302    &0.3802    &0.9998 \\
 \hline
 \multirow{2}{0.5em}{6}& P-ESCA &0.1066    &0.0662    &0.1275    &0.4123    &0.0752 \\
                          & MOFA   &0.3324    &0.3259    &0.3364    &0.3788    &0.9999 \\
 \hline
 \multirow{2}{0.5em}{7}& P-ESCA &0.0130    &0.0129    &0.0129    &0.0133    &0.0130 \\
                          & MOFA   &NA    &NA    &NA    &NA    &NA \\
  \bottomrule
\end{tabular}
\end{table}

\begin{table}[h!]
\centering
\caption*{\textbf{Tab.~S6}: Mean RMSEs in estimating the simulated parameters $\mathbf{\Theta}$, $\left\{ \mathbf{\Theta} \right\}_{l=1}^3$ and $\bm{\mu}$ derived from repeating the experiments 10 times using the P-ESCA model and the MOFA model for seven G-B-B simulation cases.}
\label{Table:S6}
\begin{tabular}{lllllll}
  \toprule
case & method & $\text{RMSE}(\mathbf{\Theta})$ & $\text{RMSE}(\mathbf{\Theta}_1)$ & $\text{RMSE}(\mathbf{\Theta}_2)$ & $\text{RMSE}(\mathbf{\Theta}_3)$ & $\text{RMSE}(\bm{\mu})$ \\
  \midrule
 \multirow{2}{0.5em}{1}& P-ESCA  &0.0376    &0.0078    &0.0463    &0.0855    &0.0210  \\
                          & MOFA  &0.1674    &0.0023    &0.3422    &0.4241    &1.0000  \\
 \hline
 \multirow{2}{0.5em}{2}& P-ESCA  &0.0415    &0.0105    &0.0544    &0.0985    &0.0167 \\
                          & MOFA  &0.1663    &0.0020    &0.3259    &0.3894    &1.0000 \\
 \hline
 \multirow{2}{0.5em}{3}& P-ESCA  &0.0552    &0.0110    &0.0708    &0.1874    &0.0231 \\
                          & MOFA  &0.2008    &0.0029    &0.3346    &0.3847    &1.0000 \\
 \hline
 \multirow{2}{0.5em}{4}& P-ESCA  &0.0712    &0.0021    &0.0986    &0.4200    &0.0750 \\
                          & MOFA  &0.1674    &0.0023    &0.3422    &0.4241    &1.0000 \\
 \hline
 \multirow{2}{0.5em}{5}& P-ESCA  &0.0775    &0.0018    &0.1107    &0.4023    &0.0806  \\
                          & MOFA  &0.1663    &0.0020    &0.3259    &0.3894    &1.0000  \\
 \hline
 \multirow{2}{0.5em}{6}& P-ESCA  &0.0731    &0.0027    &0.0878    &0.3086    &0.0626 \\
                          & MOFA  &0.2008    &0.0029    &0.3346    &0.3847    &1.0000 \\
 \hline
 \multirow{2}{0.5em}{7}& P-ESCA  &0.0107    &0.0050    &0.0129    &0.0116    &0.0107 \\
                          & MOFA  & NA       & NA       & NA       & NA       & NA \\
  \bottomrule
\end{tabular}
\end{table}

\begin{table}[h!]
\centering
\caption*{\textbf{Tab.~S7}: Mean RMSEs in estimating the simulated parameters $\mathbf{\Theta}$, $\left\{ \mathbf{\Theta} \right\}_{l=1}^3$ and $\bm{\mu}$ derived from repeating the experiments 10 times using the P-ESCA model and the MOFA model for seven G-G-B simulation cases.}
\label{Table:S7}
\begin{tabular}{lllllll}
  \toprule
case & method & $\text{RMSE}(\mathbf{\Theta})$ & $\text{RMSE}(\mathbf{\Theta}_1)$ & $\text{RMSE}(\mathbf{\Theta}_2)$ & $\text{RMSE}(\mathbf{\Theta}_3)$ & $\text{RMSE}(\bm{\mu})$ \\
  \midrule
 \multirow{2}{0.5em}{1}& P-ESCA &0.0143    &0.0089    &0.0069    &0.0555    &0.0092  \\
                          & MOFA   &0.0831    &0.0091    &0.0071    &0.5825    &1.0000  \\
 \hline
 \multirow{2}{0.5em}{2}& P-ESCA  &0.0266    &0.0126    &0.0136    &0.0955    &0.0091 \\
                          & MOFA   &0.1314    &0.0129    &0.0139    &0.7284    &1.0000 \\
  \hline
 \multirow{2}{0.5em}{3}& P-ESCA  &0.0268    &0.0137    &0.0143    &0.1158    &0.0095 \\
                          & MOFA    &0.0973    &0.0139    &0.0145    &0.6701    &1.0000 \\
 \hline
 \multirow{2}{0.5em}{4}& P-ESCA &0.0149    &0.0030    &0.0032    &0.1505    &0.0233 \\
                          & MOFA   &0.0381    &0.0031    &0.0032    &0.4400    &1.0000 \\
 \hline
 \multirow{2}{0.5em}{5}& P-ESCA &0.0174    &0.0025    &0.0025    &0.1897    &0.0314  \\
                          & MOFA   &0.0359    &0.0025    &0.0026    &0.4197    &1.0000  \\
 \hline
 \multirow{2}{0.5em}{6}& P-ESCA &0.0249    &0.0032    &0.0033    &0.1719    &0.0281 \\
                          & MOFA   &0.0527    &0.0033    &0.0034    &0.3869    &1.0000 \\
 \hline
 \multirow{2}{0.5em}{7}& P-ESCA &0.0068    &0.0050    &0.0048    &0.0123    &0.0068 \\
                          & MOFA   &NA    &NA    &NA    &NA    &NA \\
  \bottomrule
\end{tabular}
\end{table}

\begin{table}[h!]
\centering
\caption*{\textbf{Tab.~S8}: Rank estimations of the CLL data sets. Drug: drug response data; meth: DNA methylation data; mRNA: transcriptome data; mut: mutation data.}
\label{Table:S8}
\begin{tabular}{llllll}
  \toprule
  data set & data type    & size              & $k=1$ & $k=2$ & $k=3$ \\
  \midrule
  drug     & quantitative & $184 \times 310$  & 17 & 17 & 18   \\
  meth     & quantitative & $196 \times 4248$ & 8  & 9  & 9   \\
  mRNA     & quantitative & $136 \times 5000$ & 16 & 18 & 17   \\
  mut      & binary       & $200 \times 69$   & 1  & 1  & 0   \\
  \bottomrule
\end{tabular}
\end{table}

\newpage

\begin{figure}[h!]
    \centering
    \includegraphics[width=\textwidth]{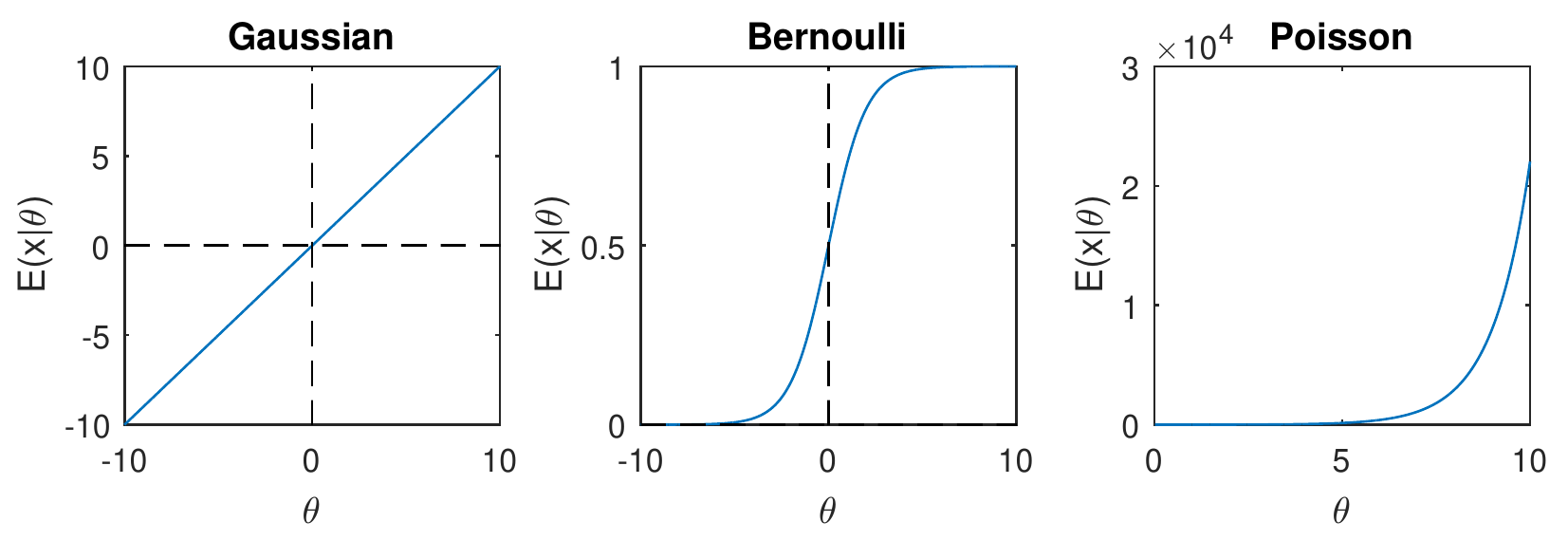}
    \caption*{\textbf{Fig.~S1}: The conditional mean of $x$, $\text{E}(x|\theta)$, for varying $\theta$ values for Gaussian, Bernoulli, Poisson distributions}
	\label{Fig:S1}
\end{figure}

\begin{figure}[h]
    \centering
    \includegraphics[width=\textwidth]{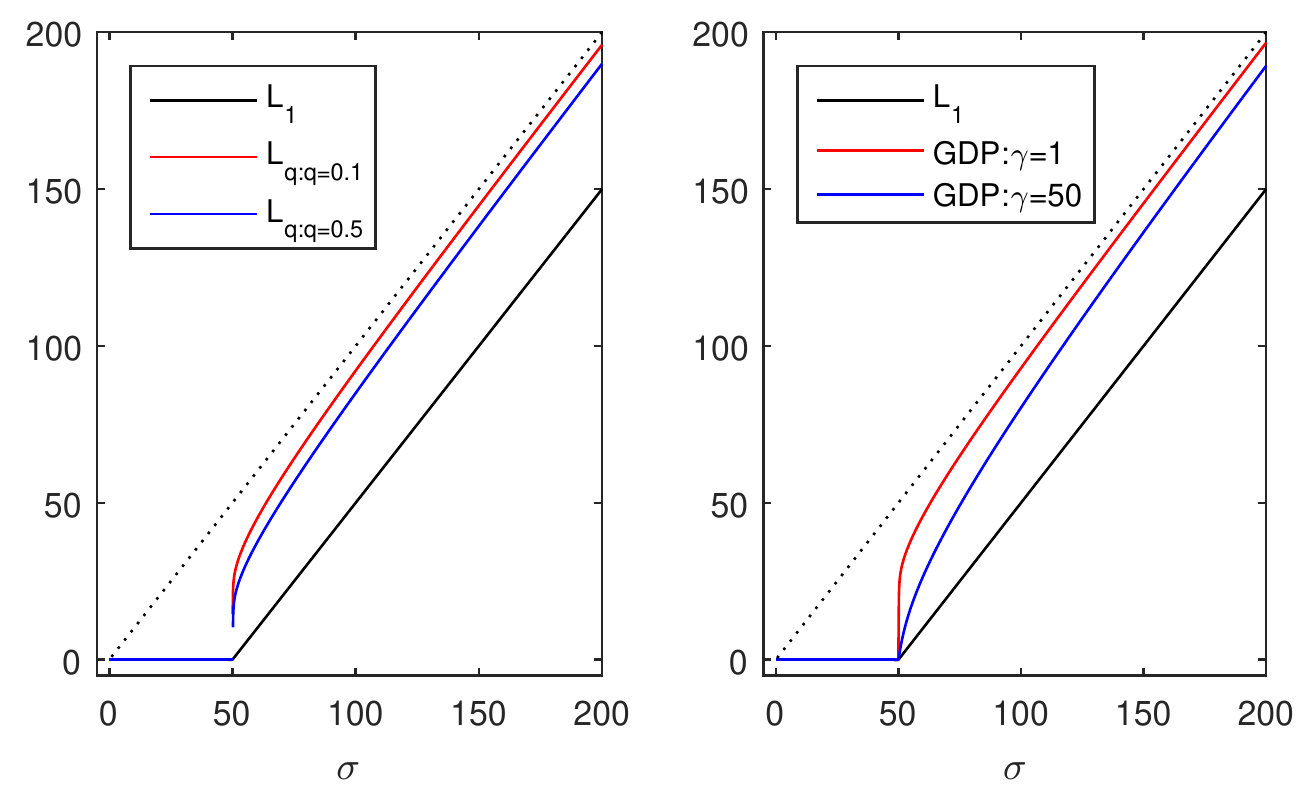}
    \caption*{\textbf{Fig.~S2}: The thresholding properties of the group lasso ($\text{L}_1$), the group $\text{L}_{q}$ and the group GDP penalty. $\sigma$ is taken as the $L_2$ norm of a group of elements. $q$ and $\gamma$ are the hyper-parameters of the corresponding penalties. $x$ axis indicates the value of $\sigma$ before thresholding; $y$ axis indicates the value after threhsolding. $\text{L}_{q: 0<q < 1}$ penalty is non-differentiable at 0.}
	\label{Fig:S2}
\end{figure}

\begin{figure}[h!]
    \centering
    \includegraphics[width=\textwidth]{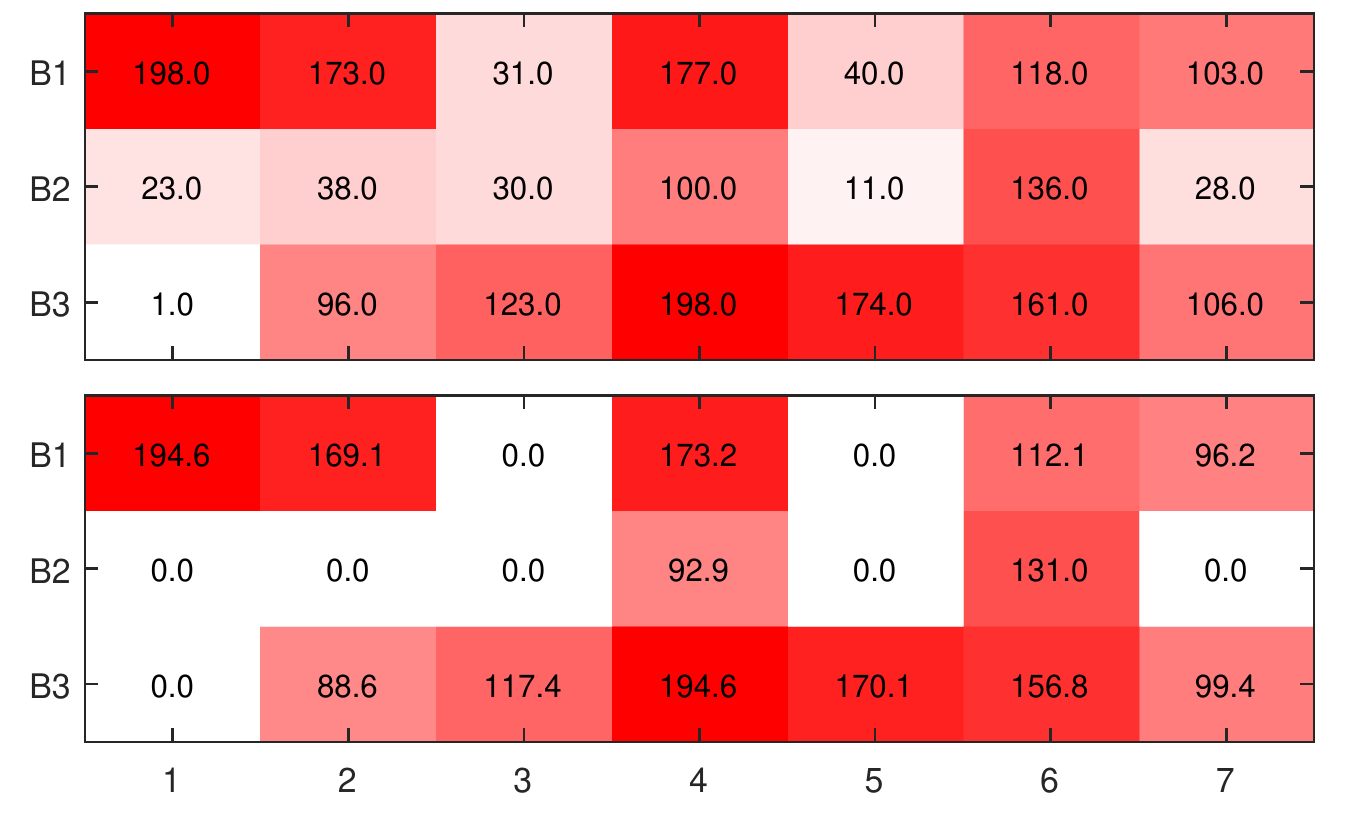}
    \caption*{\textbf{Fig.~S3}: How the group GDP ($\gamma = 1$) penalty induces structured sparse pattern on $\left\{\mathbf{B} \right\}_{l=1}^3$. Values inside the plot indicate the $\text{L}_2$ norm of the corresponding loading vector $\mathbf{b}_{l,r}$. Top: loading matrix before thresholding; bottom: loading matrix after thresholding.}
	\label{Fig:S3}
\end{figure}

\begin{figure}[h!]
    \centering
    \includegraphics[width=\textwidth]{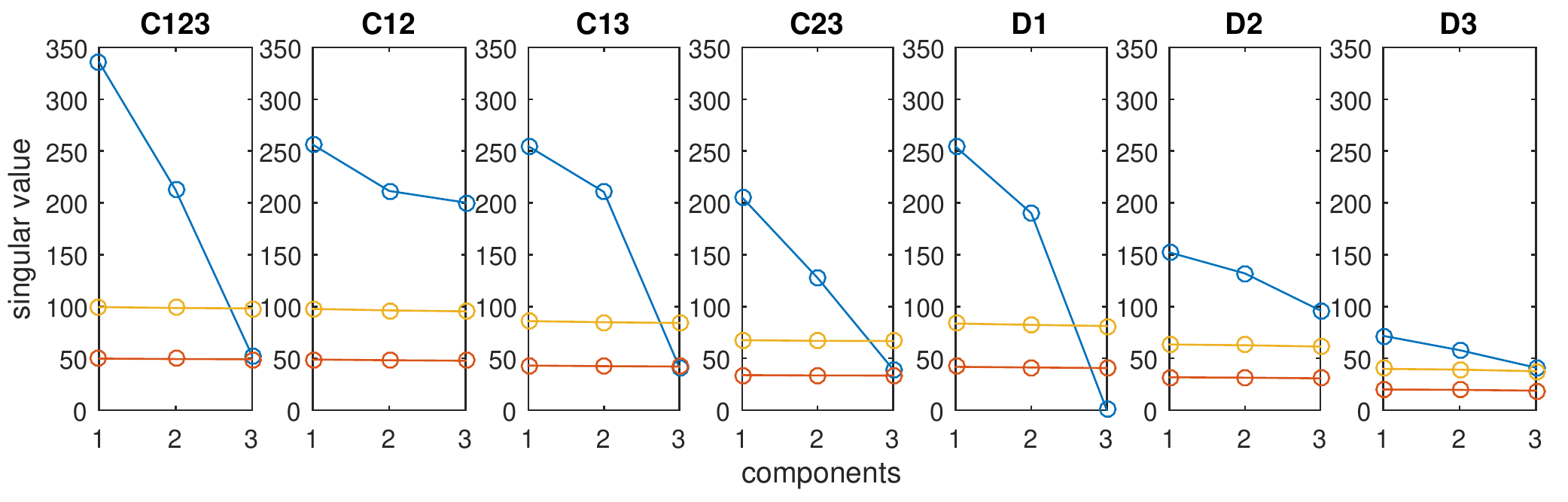}
    \caption*{\textbf{Fig.~S4}: The singular values of the simulated common, local common and distinct structures and of the corresponding residual terms. Blue dots: singular values of the simulated structure; red dots: singular values of the residual term; yellow dots: 2 times of the singular values of the residual term.}
	\label{Fig:S4}
\end{figure}

\begin{figure}[h!]
    \centering
    \includegraphics[width=\textwidth]{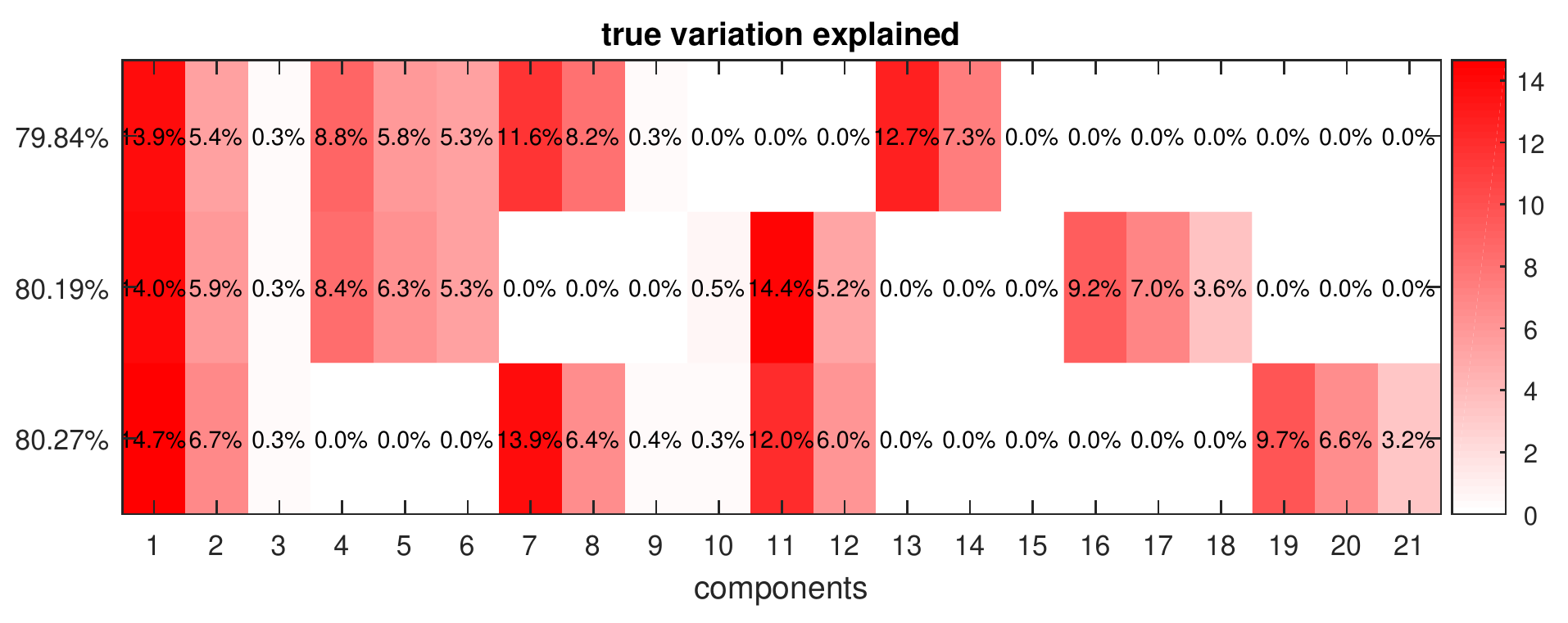}
    \caption*{\textbf{Fig.~S5}: The variation explained ratios computed using the simulated parameters. From the top to the bottom, we have data sets $\mathbf{X}_1$, $\mathbf{X}_2$ and $\mathbf{X}_3$; from the left to the right, we have 21 components corresponding to the global, local common and distinct structures. The total variation explained ratios for each data set are shown in the left of the plot, while the variation explained ratio of each component for each data set is shown inside the plot.}
	\label{Fig:S5}
\end{figure}

\begin{figure}[h!]
    \centering
    \includegraphics[width=\textwidth]{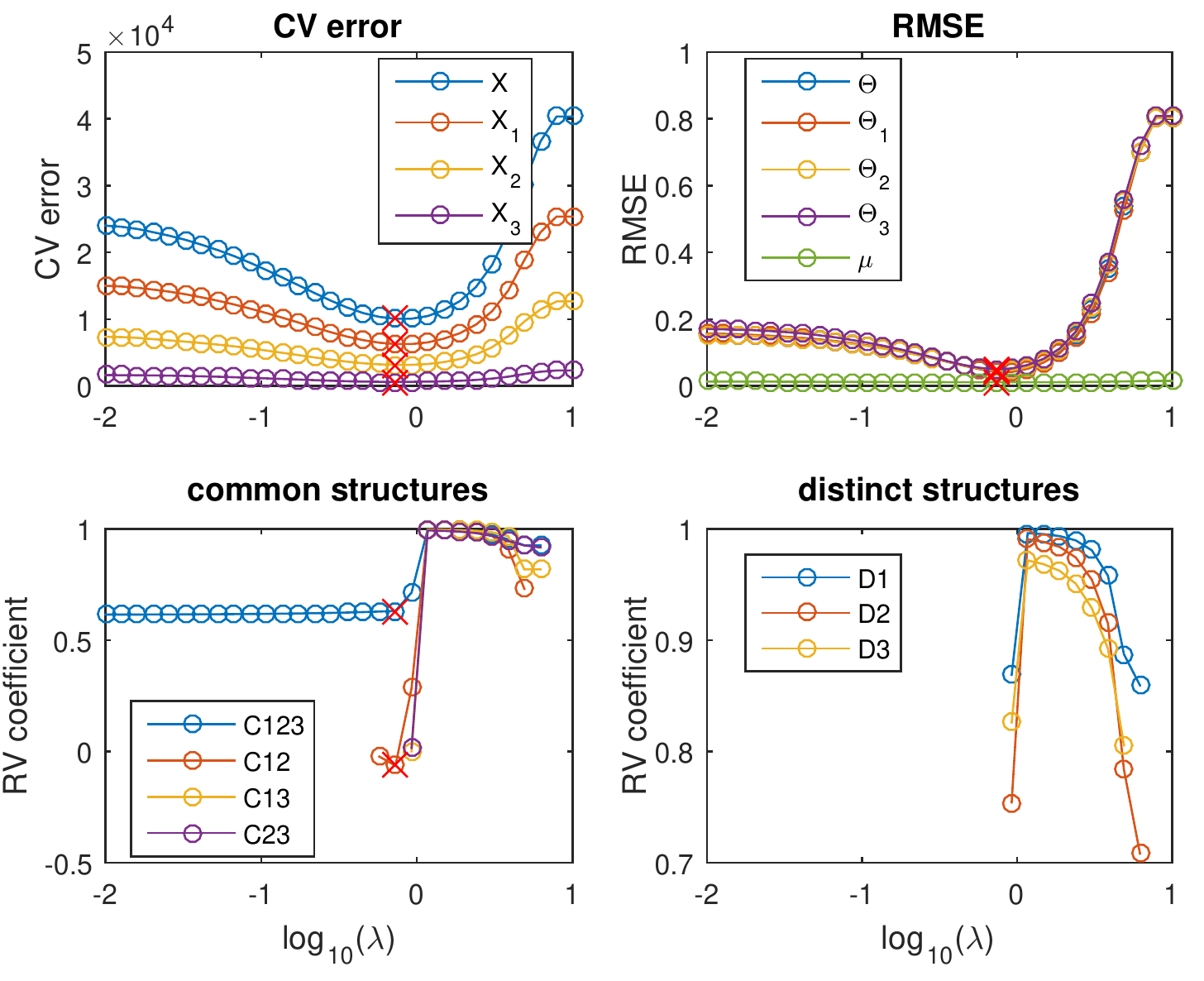}
    \caption*{\textbf{Fig.~S6}: CV errors (top left), RMSEs (top right) and the RV coefficients in estimating the common structures (bottom left), and distinct structures (bottom right) as a function of the regularization strength $\lambda$ when the P-ESCA model with a group lasso penalty is used. The red cross marker indicates the point corresponding to the minimum CV error.}
	\label{Fig:S6}
\end{figure}

\begin{figure}[h!]
    \centering
    \includegraphics[width=\textwidth]{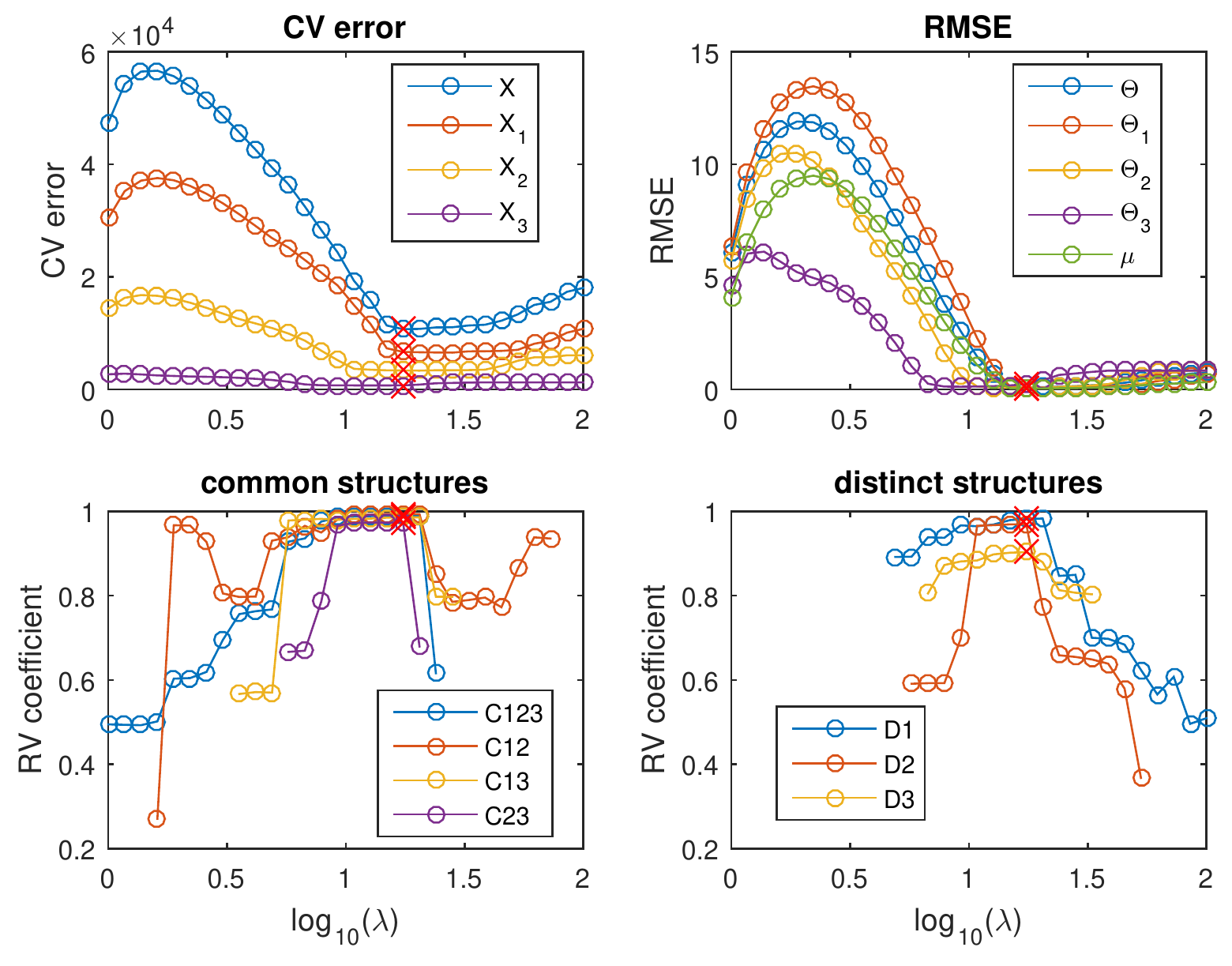}
    \caption*{\textbf{Fig.~S7}: CV errors (top left), RMSEs (top right) and the RV coefficients in estimating the common structures (bottom left), and distinct structures (bottom right) as a function of the regularization strength $\lambda$ for the P-ESCA model with a group GDP penalty on the simulated B-B-B data sets. The red cross marker indicates the point corresponding to the minimum CV error. The SNRs of global, local common and distinct structures in the B-B-B simulation are set to be 1. The reason for the increased CV errors at the early stage (top left) is that these models have not convergenced in 500 iterations.}
	\label{Fig:S7}
\end{figure}

\begin{figure}[h!]
    \centering
    \includegraphics[width=\textwidth]{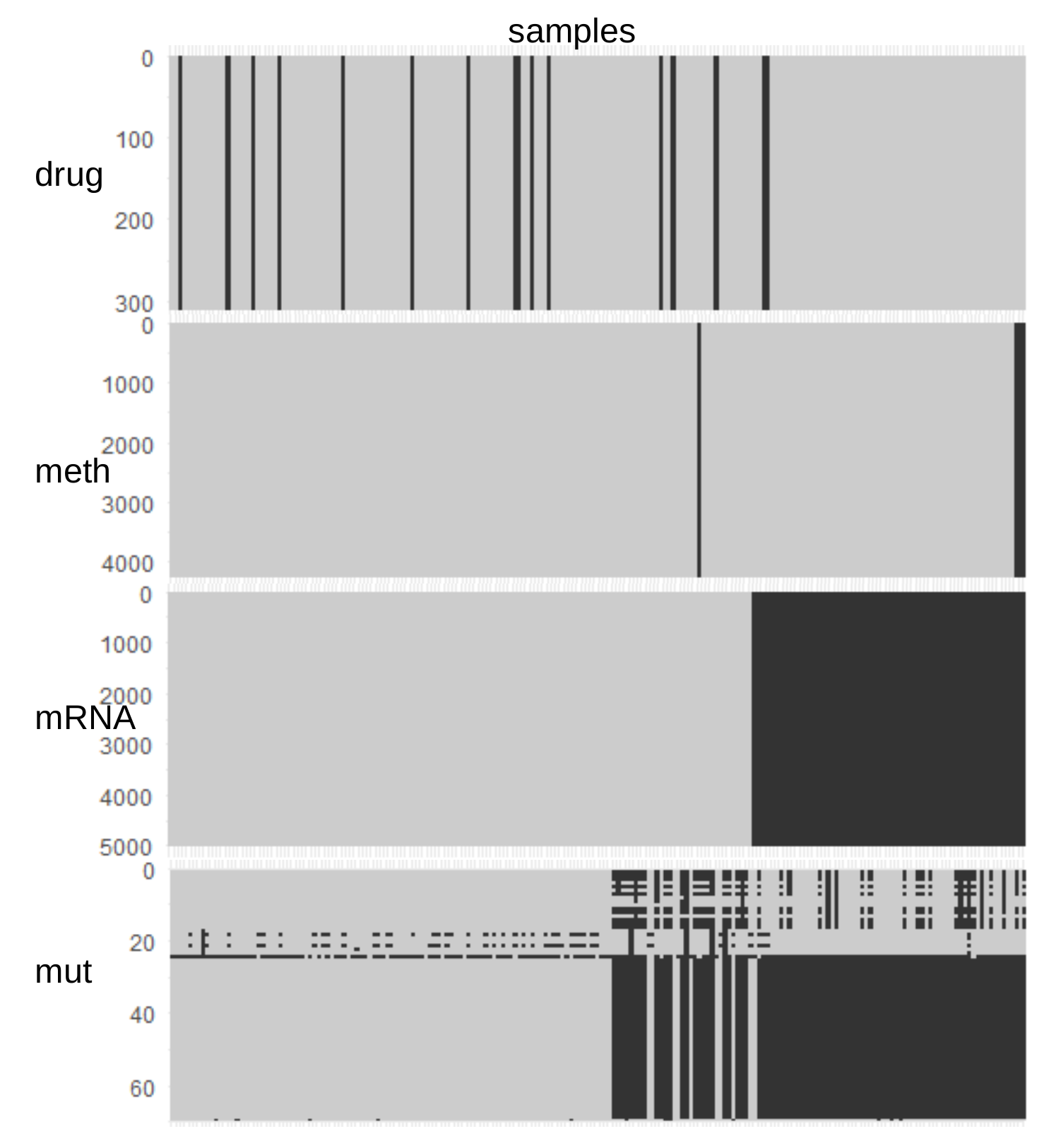}
    \caption*{\textbf{Fig.~S8}: Missing pattern of the CLL data sets. Black color indicates the data is missing, while gray color, the data is present. Drug: drug response data; meth: DNA methylation data; mRNA: transcriptome data; mut: mutation data.}
	\label{Fig:S8}
\end{figure}

\begin{figure}[h!]
    \centering
    \includegraphics[width=\textwidth]{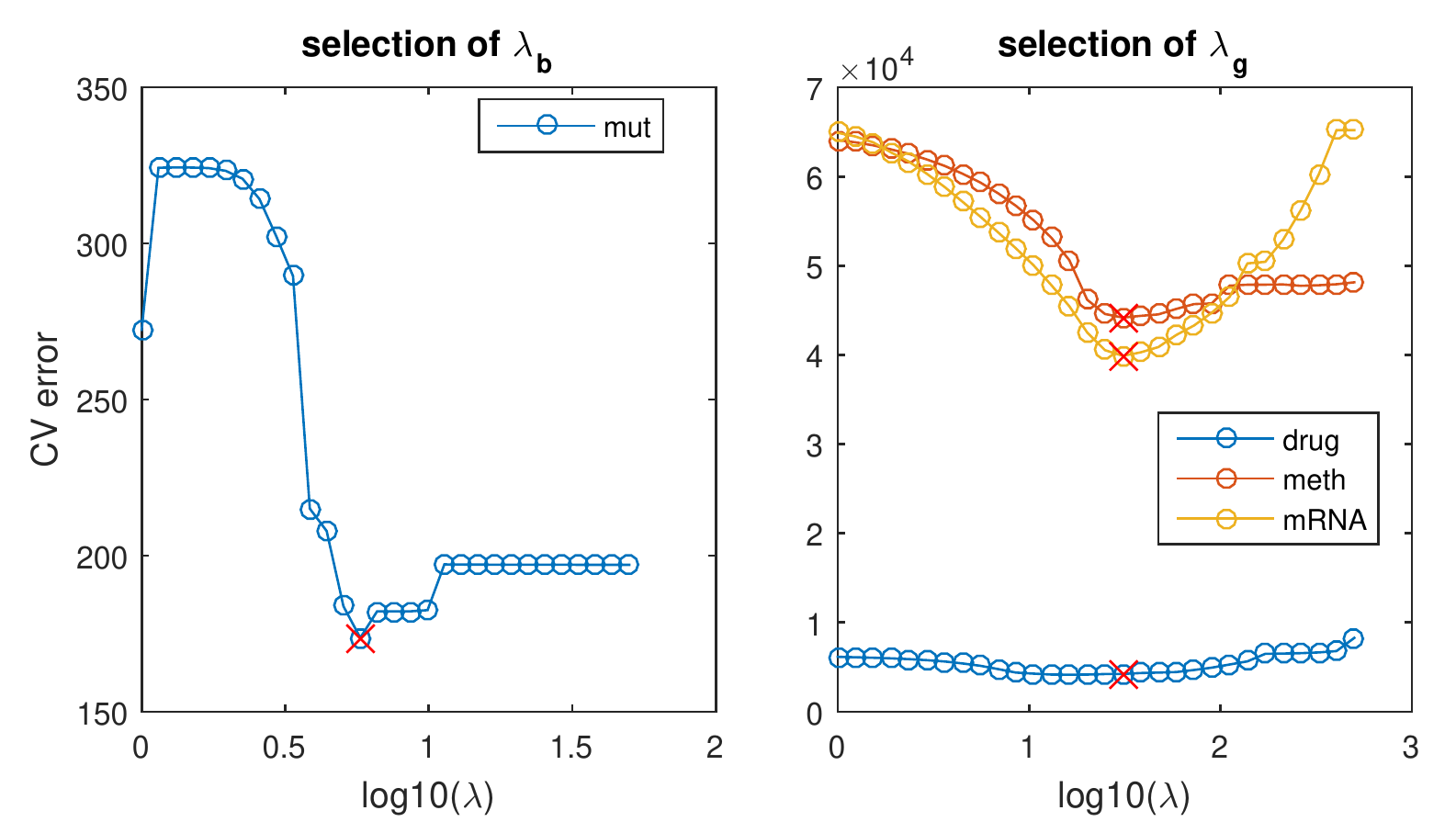}
    \caption*{\textbf{Fig.~S9}: Model selection of the P-ESCA model with a group GDP penalty on the CLL data sets. Drug: drug response data; meth: DNA methylation data; mRNA: transcriptome data; mut: mutation data. The red cross marker indicates the selected model.}
	\label{Fig:S9}
\end{figure}

\begin{figure}[h!]
    \centering
    \includegraphics[width=\textwidth]{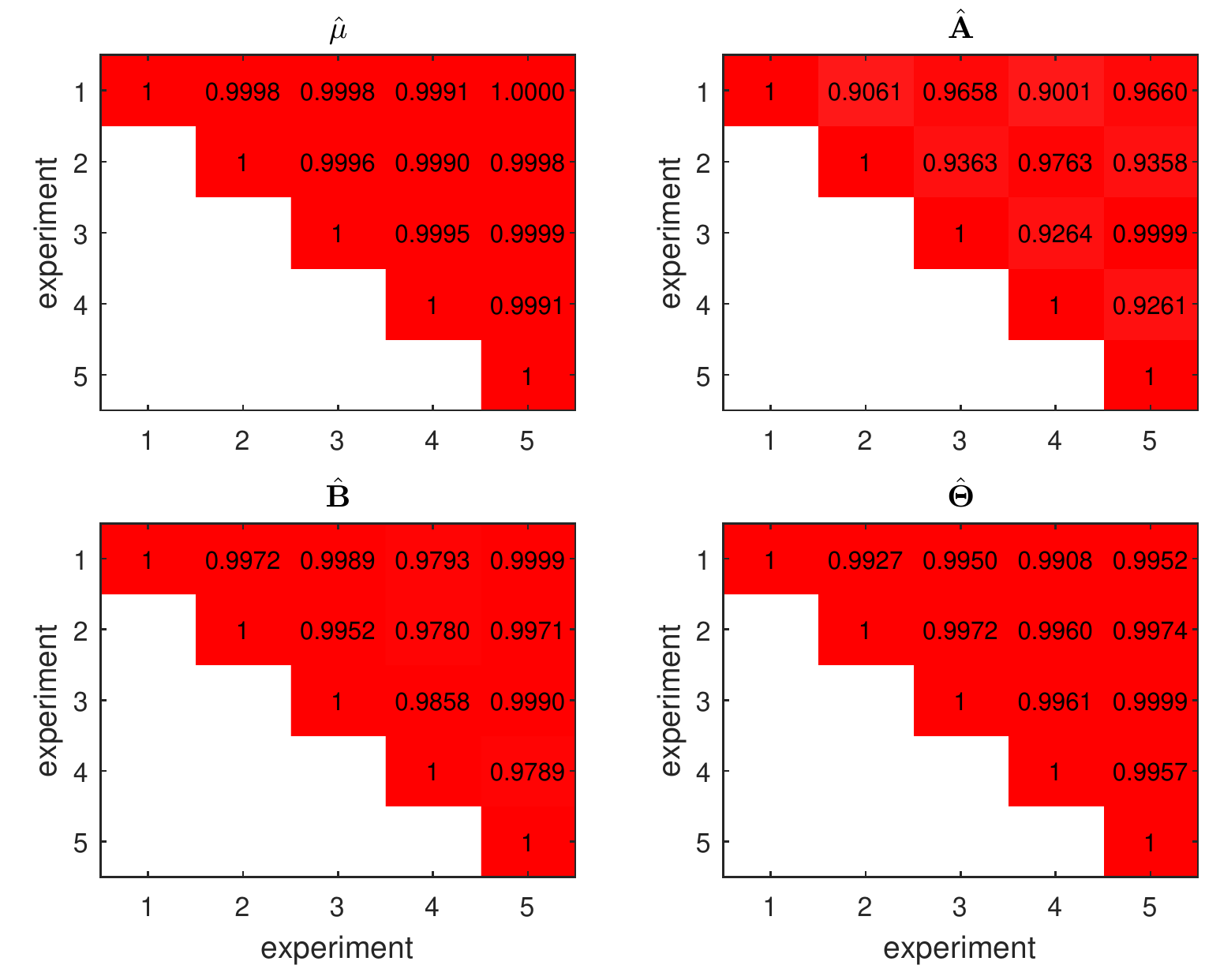}
    \caption*{\textbf{Fig.~S10}: The Pearson coefficient matrix for the 5 estimations of the $\hat{\bm{\mu}}$ and the RV coefficient matrices for the 5 estimations of the $\hat{\mathbf{A}}$, $\hat{\mathbf{B}}$ and $\hat{\mathbf{\Theta}}$ derived from the P-ESCA model.}
	\label{Fig:S10}
\end{figure}

\begin{figure}[h!]
    \centering
    \includegraphics[width=\textwidth]{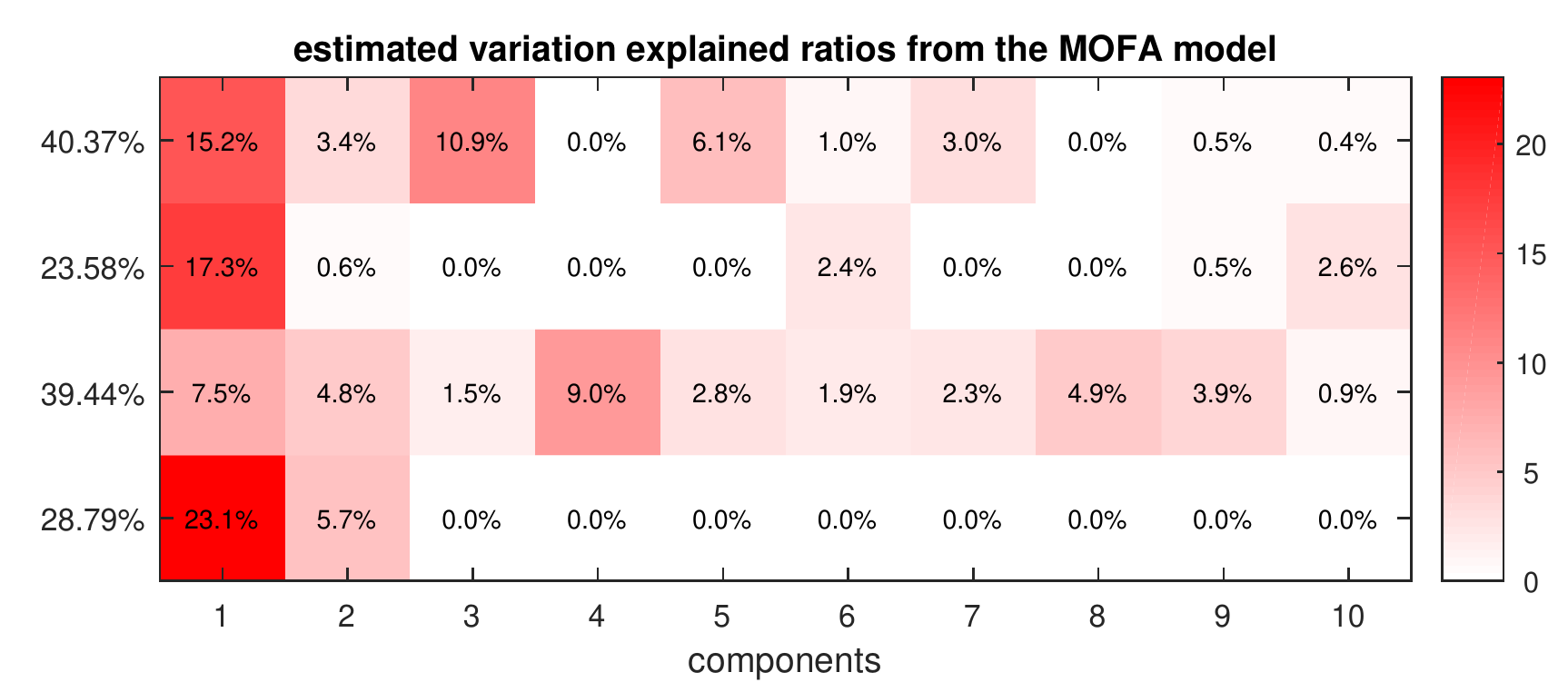}
    \caption*{\textbf{Fig.~S11}: Variation explained plots reproduced from the MOFA paper. From the top to the bottom, the data sets are drug response, methylation, mRNA and mutation}
	\label{Fig:S11}
\end{figure}

\end{document}